\documentclass[%
 reprint,
 prx,
 floatfix,
superscriptaddress,
 amsmath,amssymb,
 aps,
]{revtex4-2}

\usepackage{graphicx}
\usepackage{dcolumn}
\usepackage{bm}
\usepackage{amsmath}
\usepackage{booktabs}
\usepackage[table]{xcolor}
\usepackage{array}
\usepackage{rotating}
\usepackage{threeparttable}
\usepackage{float}
\usepackage{afterpage}

\definecolor{passgreen}{RGB}{180, 220, 180}
\definecolor{failred}{RGB}{220, 180, 180}
\definecolor{partialyellow}{RGB}{255, 235, 180}

\newcommand{\passmark}{\cellcolor{passgreen}$\checkmark$}
\newcommand{\failmark}{\cellcolor{failred}$\times$}
\newcommand{\partialmark}{\cellcolor{partialyellow}$(\checkmark)$}
\usepackage{hyperref} 

\begin{document}

\title{Long-Range Machine Learning Interatomic Potentials for Defect Energetics in SrTiO$_3$}

\author{Yee Chit Wong}
\affiliation{Department of Physics, University of Warwick, Coventry, CV4 7AL, UK}

\author{Marcel F. Langer}
\affiliation{Laboratory of Computational Science and Modeling, École Polytechnique Fédérale de Lausanne, 1015 Lausanne, Switzerland}
\author{Reinhard J. Maurer}%
\affiliation{University of Vienna, Faculty of Physics, Kolingasse 14{--}16, 1090 Vienna, Austria}
\affiliation{Institute of Physical Chemistry, Georg-August University, Göttingen 37077, Germany}
\affiliation{Max-Planck-Institute for Multidisciplinary Sciences, Göttingen 37077, Germany}
\author{Nicholas D. M. Hine}%
\email{n.d.m.hine@warwick.ac.uk}
\affiliation{Department of Physics, University of Warwick, Coventry, CV4 7AL, UK}

\begin{abstract}
Machine learning interatomic potentials (MLIPs) have advanced rapidly in recent years, yet the majority of models remain semilocal in nature and neglect long-range electrostatic interactions. A growing number of long-range models have emerged to address this limitation, but systematic benchmarks comparing their performance on physically realistic systems remain scarce. In this work, we assess three long-range MLIPs -- MACE-POLAR, MACE-LES, and LOREM, against a short-range baseline MACE model, on their ability to capture long-range Coulombic interactions, progressing from idealised synthetic point-charge systems to first-principles data on realistic crystal structures. This progression allows us to isolate model's capability to learn electrostatic interactions from the many competing energy contributions, such as dielectric screening and exchange-correlation, that are present in DFT total energies. As a physically motivated benchmark system, we focus on strontium titanate SrTiO$_3$ with Sr--O Schottky vacancy pairs, a material in which long-range charge interactions play a central role and which serves as a natural stepping stone towards the study of charged polarons and other electronic and optical properties. Building on this benchmark, we further examine oxygen vacancy migration in the presence of the Sr--O Schottky pair, using climbing-image nudged elastic band calculations and molecular dynamics to obtain activation barriers as a function of the vacancy pair separation, providing further extrapolation test of each model's ability to capture the electrostatic effects that govern defect transport.  

\end{abstract}

\maketitle
\section{Introduction} \label{sec:introduction}

The past decade has seen a rapid proliferation of machine learning interatomic potentials (MLIPs), with an increasing number of models targeting bulk materials, as well as heading toward general-purpose foundation models capable of describing diverse chemical environments \cite{batatia_foundation_2023, barroso-luque_open_2026, yang_mattersim_2024, gharakhanyan_open_2026, deng_chgnet_2023}. The foundational architecture underlying most modern MLIPs traces back to the seminal work of Behler and Parrinello~\cite{behler_generalized_2007}, who demonstrated that a high-dimensional potential-energy surface can be decomposed into a sum of local atomic energies, each inferred from a symmetry-function representation of the local chemical environment. This atom-decomposition ansatz was subsequently inherited by message-passing neural network (MPNN) architectures, in which atomic energies $E_i$ are accumulated into a total energy $E=\sum_i{E_i}$ after iterative rounds of neighbourhood aggregation. Despite this progress, a persistent limitation of many state-of-the-art MLIPs is their lack of explicit treatment of long-range interactions. Models such as PhysNet~\cite{unke_physnet_2019} and SpookyNet~\cite{unke_spookynet_2021} took an important step in this direction by predicting environment-dependent atomic partial charges, constraining them to reproduce the total system charge, and summing the resulting Coulombic interactions to capture long-range electrostatic contributions. This approach is physically well-motivated: the charge-constrained Coulomb interaction has the correct asymptotic form, and no arbitrary charge partitioning scheme need be imposed on the training data. However, these models were developed before the advent of fully equivariant message-passing architectures, and were predominantly validated on molecular systems rather than bulk polar or ionic materials, where a well-defined molecular dipole moment cannot be used to supervise charge learning. Hence, extending a rigorous electrostatic treatment to the more expressive, high-body-order backbones of modern MLIPs remains a key challenge, particularly for ionic and polar crystalline systems where Coulombic interactions decay slowly and shape the potential energy landscape.

Recent efforts have begun to address this deficiency through various approaches. Attention-based mechanisms, as in RANGE, introduce additional broadcasting and aggregation layers during message passing to extend the effective interaction range \cite{caruso_extending_2025}; however, this is most naturally suited to aperiodic systems such as long molecular chains, and its extension to periodic boundary conditions remains conceptually unclear. Latent Ewald summation, as employed in MACE-LES, maps atomic features into a learnable latent charge space and performs electrostatic summation directly on these latent features, sidestepping the need for physical charges \cite{cheng_latent_2025}. LOREM instead evaluates the Madelung potential explicitly via particle-mesh Ewald, encoding the long-range contribution through equivariant charge embeddings that are iteratively refined during message passing \cite{rumiantsev_learning_2025}. MACE-POLAR augments the MACE architecture with explicit atomic multipoles and performs a charge equilibration step following short-range message passing, before evaluating a reciprocal-space summation over smeared multipole densities. This equilibration enforces exact conservation of the total charge and spin, and gives the charge redistribution a physical basis through learned Fukui functions, despite the charges not being self-consistent \cite{batatia_mace-polar-1_2026}. A related approach is NequIP-LR, which adopts the charge equilibration scheme of fourth-generation high-dimensional neural network potentials within an equivariant architecture \cite{ko_fourth-generation_2021, maruf_learning_2025}. Rather than predicting charges directly, a network predicts atomic electronegativities, from which scalar partial charges are obtained by minimising a charge-dependent energy expression under a fixed total charge. The resulting charges enter both an explicit Coulomb term and the short-range energy model. The diversity of these approaches reflects both the importance and the open nature of the problem. 

In this work, we seek to understand the performance of MLIPs for defects in oxide materials, using strontium titanate (SrTiO$_3$) as a prototypical example. As a perovskite oxide, SrTiO$_3$ is a natural testing ground for this purpose: long-range electrostatics play a central role in its defect physics \cite{de_souza_formation_2009, yang_determination_2021, mott_conduction_1938, Zhuang_srtio3surfaces_2014, zhang_oxygen_2016}. Especially for Schottky defect formation, where pairs of charged vacancy sites give rise to spatially localised charge densities in an overall charge-neutral supercell, it provides a well-defined and physically transparent benchmark. Understanding defect-defect interactions in multiscale simulations of such systems demands a model whose energy and force predictions generalise reliably beyond the length scales represented in the training data, which leads to a stringent test of long-range transferability \cite{freysoldt_first-principles_2014, hooven_how_2026}. Furthermore, although density functional theory (DFT) is deterministic in principle, practical calculations are subject to numerical errors arising from incomplete basis sets and finite SCF convergence, which limits the precision of the reference labels \cite{carbogno_numerical_2022, reproducibility_2016, kresse_efficiency_1996, woods_computing_2019}. Since long-range electrostatic contributions to the energy and forces can be small, such errors may obscure whether a model has genuinely learned the correct physics. It is therefore necessary to first establish whether current long-range MLIPs can faithfully recover idealised electrostatic behaviour, before deploying them on \textit{ab initio} datasets. 

In this study, we evaluate and compare several MLIPs with explicit long-range interaction, namely MACE-LES, MACE-POLAR, and LOREM, alongside a short-range MACE model as a baseline. We assess their suitability for characterising our target systems. The study proceeds in two stages. We first construct synthetic training sets with analytically exact Ewald energies and forces for point-charge systems and synthetic SrTi$\text{O}_3$ structures. Then we use the synthetic datasets to benchmark the ability of leading long-range MLIP architectures to learn and extrapolate Coulombic interactions in a controlled setting. Having established where each model succeeds and where fundamental limitations remain, we then evaluate the same models on DFT reference data for SrTiO$_3$, allowing us to assess how well the capabilities identified in the synthetic regime translate to the complexity of the real \textit{ab initio} reference data. Together, these two stages provide a comprehensive picture of the practical readiness of current long-range MLIPs for polar oxide systems. Finally, we use the models to investigate oxygen vacancy migration and diffusion. Computational studies on diffusion properties are generally vulnerable to finite-size errors and the limited timescales accessible to \textit{ab initio} molecular dynamics (AIMD) \cite{yeh_system-size_2004, grasselli_investigating_2022, celebi_finite-size_2021, marx_ab_2000}. The long-range MLIPs developed here offer a practical alternative to overcome both limitations by enabling large-scale simulations over extended trajectories at a fraction of the computational cost.

\section{Theory} \label{sec:theory}

All MLIPs considered in this study share a common graph neural network backbone: the total energy is decomposed as the energy contribution from all individual nodes, $E= E(\{\mathbf{r}_i, Z_i\}) = \sum_i{E_i}$ \cite{schutt_quantum-chemical_2017, unke_physnet_2019}. The short-range component is broadly consistent across models, and is computed via iterative message passing over a local neighbourhood graph, where each node aggregates features from neighbouring atoms within a fixed cutoff radius (typically 5-8~\AA), updating a feature vector $\mathbf{h}_i$ \cite{behler_generalized_2007,batzner_e3eq_2022,Bochkarev_grace_2024}. The message-passing step can consist of multiple layers to extend the effective receptive field indirectly. With $T$ layers of cutoff radius $r_c$, information can in principle propagate up to $T \times r_c$. For example, a model with two layers of 5~\AA\ has an effective receptive field of up to 10~\AA. However, this transmission is mediated through intermediate atoms along the graph, so the actual information range depends on the local geometry \cite{batatia_design_2025}. Atoms must be connected by a chain of neighbours within $r_c$, and the effective range can be shorter in sparse or low-coordination environments. The feature vector is generally denoted as $\mathbf{h}_i$, with $i$ representing the atom index. All models considered employ both invariant and equivariant feature representations. The equivariant features transform as irreducible representations of SO(3), indexed by degree $l$, where $l=0$ recovers the invariant (scalar) channel and higher degrees $l>0$ capture directional, symmetry-constrained information and transform according to irreducible representations under rotations. The number of features per degree $l$ is referred to as the channel multiplicity\cite{thomas_tensor_2018,batatia_mace_2022}.  

\subsection{MACE-LES}
MACE-LES comprises an additive long-ranged model $E_{\mathrm{LR}}$ added to the short-range MACE model $E_{\mathrm{SR}}$ for a total energy $E = E_{\mathrm{SR}} + E_{\mathrm{LR}}$ \cite{cheng_latent_2025, kosmala_ewald-based_2023, grisafi_incorporating_2019}. To calculate the long-range energy, the latent charges $q_i^{\mathrm{les}}$ are predicted directly from the feature vector $\mathbf{h}_{i}$ through a multilayer perceptron (MLP) network, i.e., $q_i^{\mathrm{les}} = \text{MLP}(\mathbf{h}_i)$. The long-range energy is simply the reciprocal space term of the standard form of the Ewald summation:
\begin{equation} \label{eq:mace-les-lr-energy}
    E_{\mathrm{LR}} = \frac{1}{\Omega} \sum_{0<|\mathbf{k}|<k_{\mathrm{cutoff}}} {\frac{1}{k^2}e^{-\frac{\sigma^2 k^2}{2}} \bigg|\sum_i {q_i^{\mathrm{les}} e^{(-i\mathbf{k} \cdot \mathbf{r}_i)}}\bigg|^2} \;,
\end{equation}

\noindent where $\Omega$ is the volume of the cell, $\mathbf{k}$ are the reciprocal lattice vectors and $\sigma$ is a chosen Gaussian smearing width. The wavevector cutoff $k_{\mathrm{cutoff}}$ serves to exclude short-ranged interactions from $E_{\mathrm{LR}}$ and is treated as a hyperparameter to balance computation time and accuracy. In this approach, the short-range part of the Ewald summation is implicitly absorbed into $E_{\mathrm{SR}}$. One potential issue with this approach is that the latent charges are unconstrained in sign, magnitude and net sum. In a standard Ewald treatment, a non-zero net charge causes the $k=0$ reciprocal-space term to diverge. However in this formalism, this term is excluded from the sum by construction, so no divergence occurs regardless of whether the underlying charge assignment is neutral. Because this exclusion is unconditional rather than tied to a genuine neutrality constraint, $q^{\mathrm{les}}$ has no direct physical interpretation.

\subsection{MACE-POLAR}
MACE-POLAR is a model built on top of the short-range MACE model \cite{batatia_mace-polar-1_2026}. The modification mainly comes in the computation of the total energy:
\begin{equation} \label{eq:MACE-polar-energy}
    E = E_{\mathrm{local}} + E_{\mathrm{non-local}} + E_{\mathrm{elec}}.
\end{equation}

\noindent The non-local energy term $E_{\mathrm{non-local}}$ captures interactions such as dispersion and charge distribution changes with the field, while the electrostatics term $E_{\mathrm{elec}}$ is simply the Coulomb interaction between the partial charges. The key principle behind MACE-POLAR is to construct a physics-informed charge representation using Gaussian functions, given as:
\begin{equation} \label{eq:MACE-polar-gaussian-charge}
    \rho(\mathbf{r}) = \sum_{i, lm} {p_{i, lm} \, \phi_{lm}(\mathbf{r} - \mathbf{r}_i)}
\end{equation}
where 
\begin{equation} \label{eq:MACE-polar-charge-basis}
    \phi_{lm} = C_{nl}|\mathbf{r} - \mathbf{r}_i)|^l \, e^{-\frac{|\mathbf{r} - \mathbf{r}_i)|^2}{2 \sigma^2}} Y_{lm}(\mathbf{r} - \mathbf{r}_i) 
\end{equation}
are the basis functions and $p_{i, lm}$ is an equivariant tensor containing the atomic multipole coefficients of the spin-charge density expansion. $p_{i, lm}$ encodes the monopole ($l=0$), dipole ($l=1$) and higher-order multipole ($l \geq 2$) contributions, which are derived directly from the feature vector $\mathbf{h}_i$. One key difference is in the treatment of charge conservation: MACE-POLAR introduces a charge equilibration scheme that forces monopole conservation via Fukui equilibrium, where
\begin{equation} \label{eq:MACE-polar-Fukui}
    p_{i, 00} = \tilde{p}_{i, 00} + \frac{f_i}{\sum{f_i}} \Delta Q.
\end{equation}
The Fukui number $f_i$ is also derived from the local feature vector $\mathbf{h}_i$, and aims to describe the softness (e.g., electronegativity) of atoms. $\Delta Q$ is the difference between the predicted charge and the actual charge of the system. This charge redistribution scheme is conceptually related to classical electronegativity equalisation methods~\cite{mortier_electronegativity-equalization_1986, rappe_charge_1991} and their neural network generalisations~\cite{ghasemi_interatomic_2015}, in which atomic softness parameters govern how excess charge is distributed across atoms to satisfy a global constraint.  

Before the final energy contributions are computed, the initial local multipole guess $\tilde{p}^{(0)}_{i, lm}$ is refined through two iterative polarisable field updates. Each iteration proceeds in three steps: first, the electrostatic potential $v^{(u)}(\mathbf{r})$, computable from the current spin-charge density, is projected onto atom-centred basis functions to yield equivariant field features $v^{(u)}_{i,nlm}$; second, these non-local features are combined with the local node features $\mathbf{h}_i$ to predict an additive update $\Delta p^{(u)}_{i, lm}$ to the multipoles; third, the updated monopoles are re-equilibrated to enforce charge and spin conservation via a Fukui-function-based scheme inspired by conceptual DFT~\cite{parr_absolute_1983, geerlings_conceptual_2003} and first introduced into MLIPs by AIMNet-NSE~\cite{zubatyuk_teaching_2021}. This fixed-depth iterative update allows the model to capture polarisation and long-range charge transfer without the cost or potential instability of running field updates to full self-consistency. 

The non-local energy contribution captures residual long-range effects beyond pure Coulombic interactions, such as dispersion, as a learned correction parametrised jointly on local geometry and non-local electrostatic information:
\begin{equation} \label{eq:MACE-polar-non-local-energy}
E_{\text{non-local}} = \sum_i \text{MLP}(\mathbf{d}^p_i \| \mathbf{d}^v_i),
\end{equation}
where $\mathbf{d}_i^p$ and $\mathbf{d}_i^v$ are rotationally invariant scalars obtained by contracting the local node feature vector $\mathbf{h}_i$ with spin-charge multipoles $\mathbf{p}_i$ and electrostatic features $\mathbf{v}_i$ respectively, which is done by summing over the $m$ indices:
\begin{align} \label{eq:MACE-polar-non-local-contractions}
d^p_{i,lk} &= \sum_{k'} W^{p,\text{dot}}_{lkk'} \sum_{m=-l}^{l} h_{i,klm} p_{i,k'lm}, \notag \\
d^v_{i,lk} &= \sum_{k'} W^{v,\text{dot}}_{lkk'} \sum_{m=-l}^{l} h_{i,klm} v_{i,k'lm}.
\end{align}

Here $v_{i,nlm}$ are the equivariant electrostatic features obtained by projecting the spin-resolved Coulomb potential of the spin-charge density onto atom-centred Gaussian basis functions  $\phi_{lm}$. In principle, MACE-POLAR also supports an externally applied field through an additional potential $v_{\text{app}}$, but since no external field is present in our bulk periodic system, this term is not relevant to this work. Eq. \ref{eq:MACE-polar-non-local-energy} is therefore a correction that depends on the local geometry and on non-local information only through the physically motivated spin-charge density and its induced potential, a design intended to improve extrapolation beyond the training distribution.

After the iterative multipole refinement and Fukui equilibration, the monopole is then used to obtain the electrostatic energy computed in the reciprocal space using the Fourier series of the charge density $\tilde{\rho}(\mathbf{k})$:
\begin{equation} \label{eq:MACE-polar-electrostatics-energy}
    E_{\text{elec}} = \frac{1}{2 \Omega} \sum_{0<|\mathbf{k}|<k_{\mathrm{cutoff}}} {\frac{4\pi}{k^2} \bigg| \tilde{\rho}(\mathbf{k}) \bigg|^2},
\end{equation}
where $\mathbf{k}$ is the set of reciprocal lattice vectors within the cutoff. The Fourier series of charge density $\tilde{\rho}(\mathbf{k})$ can be computed analytically by:
\begin{equation} \label{eq:MACE-polar-charge-density-fourier}
    \tilde{\rho}(\mathbf{k}) = \sum_{i, lm} {p_{i, lm} \tilde{\phi}_{lm}} (\mathbf{k}) e^{-i\mathbf{k} \cdot \mathbf{r}_i} . 
\end{equation}
This expression is inherently long-range due to the computation of electrostatic energy in the reciprocal space, making the contribution dependent on the low-wavelength component. In practice, the sum is truncated at $k_{\mathrm{cutoff}}$, which in MACE-POLAR is determined adaptively from the Gaussian smearing width and the maximum angular momentum of the basis, rather than treated as an independent hyperparameter as in MACE-LES \cite{cheng_latent_2025}.

\subsection{LOREM}
One feature that sets the LOREM model apart from the others is the implementation of equivariant charges $\mathbf{Q}_j^l$, where $l$ denotes the angular momentum label \cite{rumiantsev_learning_2025}. The idea is to compute a potential that does not just depend on scalar partial charges, but instead considers the potential as an equivariant object:
\begin{equation} \label{eq:lorem-eqv-charge}
    \mathbf{V}^l_i(\mathbf{r}) = \sum_{j=1}^{N} \sum_{\mathbf{n}} {\frac{\mathbf{Q}_j^l}{{|\mathbf{r} - (\mathbf{r}_j + \mathbf{C}{\mathbf{n}})|}^p}}
\end{equation}
where $\mathbf{n}$ represents the lattice index vector enumerating the periodic cell replicas, and $\mathbf{C}$ is the lattice matrix. The conditionally convergent sum over periodic images $\mathbf{n}$ is evaluated via Ewald summation, which reformulates it into absolutely convergent real and reciprocal-space sums that can be truncated unambiguously \cite{rumiantsev_learning_2025}. $p$ is the power law exponent of the interaction kernel. In LOREM, it is set to $p=1$ (Coulomb), and it is independent of $l$. The equivariant charges $\mathbf{Q}_i^l$ are derived from the feature vector via linear projection (spherical equivariance component) and MLP (scalar component). However, there is a key architectural difference between LOREM and the other models considered here, as LOREM does not decompose the total energy into explicit short and long-range components. After Ewald summation, the resulting equivariant potentials $\mathbf{V}^l_i$ act as long-range messages: they are contracted to an invariant norm and concatenated with the model's scalar long-range potential, and this combination is used to update the feature $\mathbf{h}_i$ through a residual update block. Because this update acts directly on $\mathbf{h}_i$ rather than on a separate long-range energy term, the total energy is read out as $E=\sum_i{\mathrm{MLP}(\mathbf{h}_i)}$ after the long-range message passing step. This design encodes long-range electrostatic information implicitly within the learned representations, at the cost of interpretability: the long-range contribution cannot be isolated.

\section{Method}
To benchmark the long-range capabilities of the selected MLIP architectures, we devise a systematic framework that proceeds from idealised synthetic systems to realistic DFT data. Each stage is designed to isolate a specific aspect of long-range electrostatic behaviour, allowing us to attribute model successes and failures to well-defined physical or mathematical properties. The following subsections describe the purpose of each experiment and the procedure by which the corresponding data were generated. All models are implemented using the published versions of the models, and default hyperparameters except where stated, with a short-range cutoff of 5 $\textbf{\AA}$. The accuracy of the ML models are evaluated based on mean absolute errors (MAE) and root-mean-squared errors (RMSE).  

\subsection{Synthetic unmediated point charge system} \label{sec:method-point-charge-exp}

\begin{figure}[h!]
    \centering
    \includegraphics[width=1.0\linewidth]{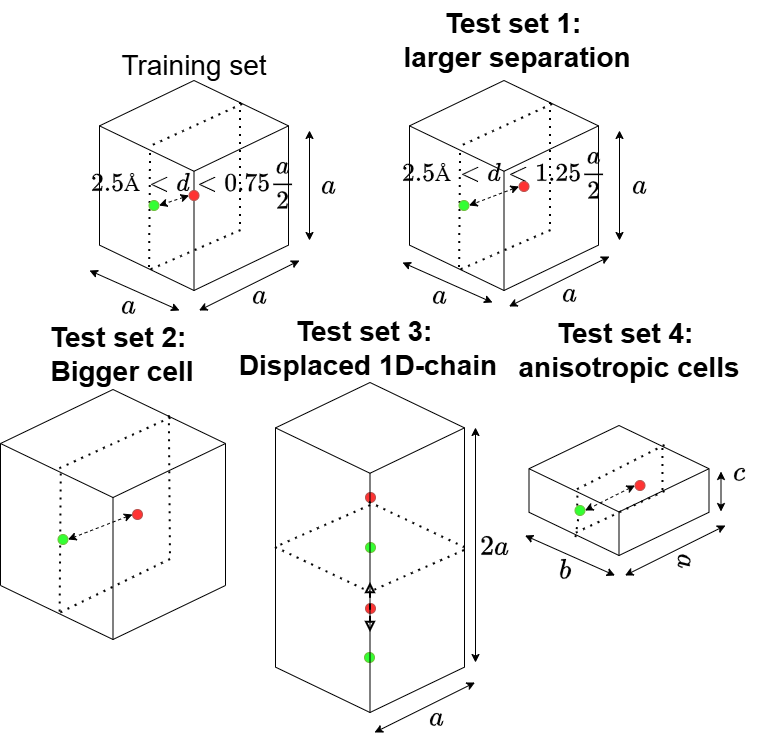}
    \caption{ Illustration of the training set and the test set for the simple point charges system. Green and red circles represent point charges of $+2e$ and $-2e$ respectively.}
    \label{fig:point_charge_dataset}
\end{figure}

As a first test, we consider a minimal synthetic system consisting of two point charges of opposite sign $q=\pm 2e$, separated by a distance $d$, placed in a large cubic periodic cell with lattice constant $a = 80$\,\AA. Ewald energies and forces are evaluated analytically using the Matscipy package \cite{grigorev_matscipy_2024}, providing noise-free reference data with exact long-range electrostatics. Training structures are restricted to separations $d < 0.375a$, with the region $0.375a \leq d \leq 0.5a$ withheld entirely for extrapolation testing as shown in Fig. ~\ref{fig:point_charge_dataset} (test set 1). This design probes a core requirement of any long-range model: the ability to generalise Coulombic interactions to length scales not represented during training.  

Beyond this basic interpolation and extrapolation test, we subject each model to three additional evaluations of increasing physical subtlety as depicted in Fig. \ref{fig:point_charge_dataset}. First, we assess size transferability by evaluating models on cells with larger lattice constants than the training cell ($a = 80$\,\AA), testing whether the learned electrostatics scale correctly with the periodic boundary conditions (test set 2). Second, we probe geometric transferability by constructing a displaced 1D-chain test by replicating the simulation cell along the $z$-direction to form a $1 \times 1 \times 2$ supercell, placing the two charges at positions $[0, d, a, \tfrac{3}{2}a]$ along $z$ (test set 3). By varying $d$, this mimics the sublattice displacement pattern of a zone-boundary phonon mode, providing a stringent test of whether models can capture the correct energy landscape under collective ionic displacements. Third, we place the same point charges in non-cubic cells to assess whether models generalise to lattice parameters and cell shapes not encountered during training (test set 4).

\subsection{Synthetic Mediated Formal Charges in Bulk SrTiO$_3$}
\label{sec:method-schottky-synthetic}

\begin{figure}[h!]
    \centering
    \includegraphics[width=1.0\linewidth]{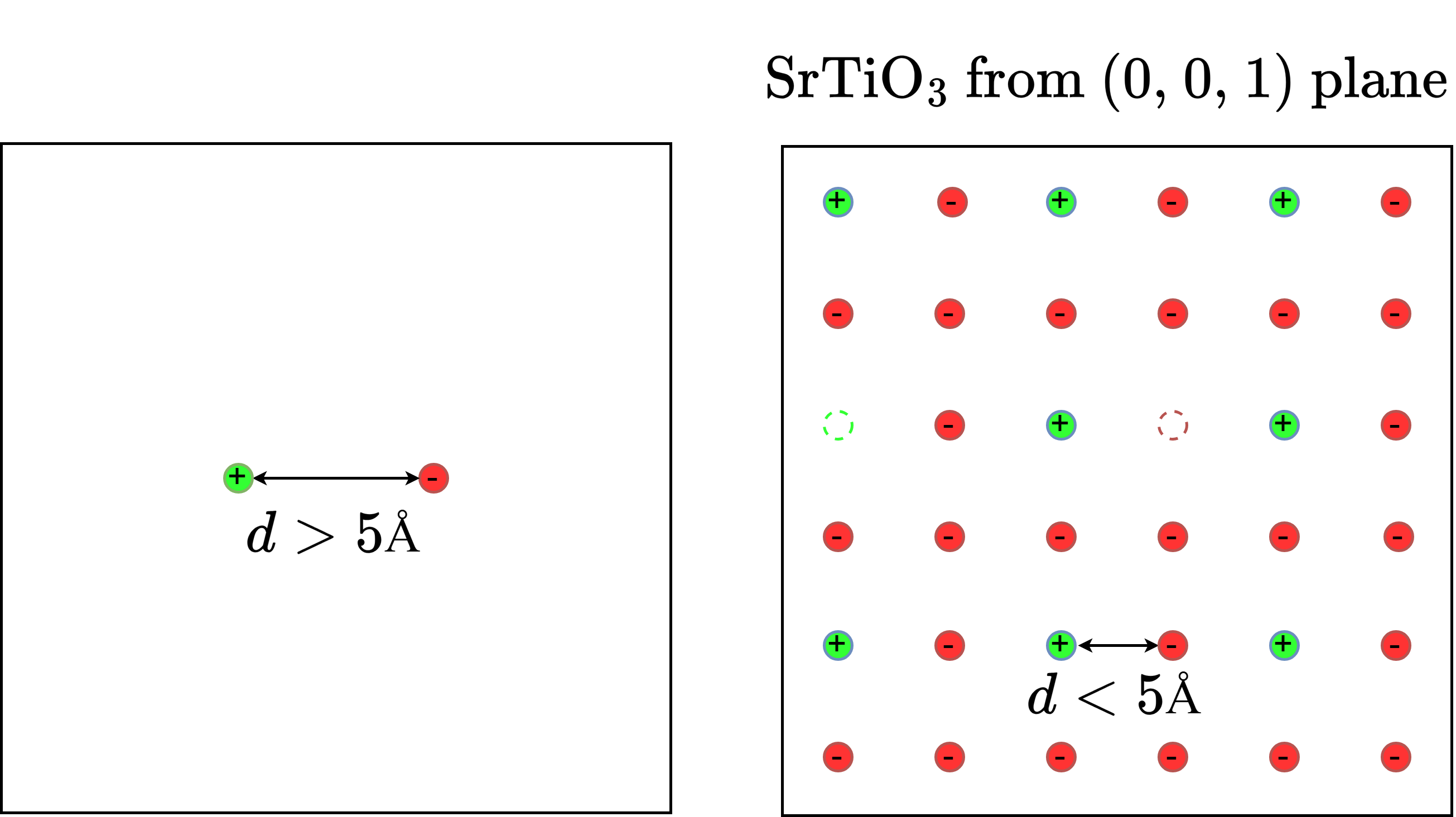}
    \caption{\label{fig:point_charge_vs_bulk} 2D schematic comparison of short-range message passing in a two-point-charge system (left) and in a bulk SrTiO$_3$ crystal looking from the (0, 0, 1) plane (right). In the bulk case, atomic mediators bridge long-range interactions across multiple message-passing hops. In the point charge system, no mediating atoms are present, and any model with a finite cutoff radius is entirely blind to interactions beyond it.}
\end{figure}

While the unmediated point charge experiment is informative, it represents an unrealistic scenario in which there are no neighbouring atoms to mediate short-range interactions or screening effects. In real bulk crystals, each atom is embedded within its neighbours. This raises the question of to what extent short-range message passing can implicitly capture long-range electrostatic effects through multi-hop atomic mediations, a question that is especially pertinent for the bare MACE model, which lacks any explicit long-range component. As illustrated in Fig. \ref{fig:point_charge_vs_bulk}, once the charge separation exceeds the cutoff radius, the two point charges become mutually invisible to a short-range model, as there exists no intermediate atom to relay information across the gap. In bulk crystals, the same model may still perform reasonably well by propagating electrostatic information through chains of neighbouring atoms due to additional layers of message passing. Consequently, strong performance on the point charge benchmark is a necessary but not sufficient condition for a model to be considered truly long-range capable, and poor performance there may overstate the practical limitations of short-range models in realistic bulk environments.  

In this part of the study, we consider SrTiO$_3$ in the cubic perovskite phase as our bulk test system, featuring charges of $+2e$ on Sr, $+4e$ on Ti, and $-2e$ on O, based on the formal oxidation states of each element in this system \cite{ghosez_dynamical_1998}. This system represents a fully ionic oxide in which long-range Coulomb interactions are substantial and must be accurately reproduced throughout the periodic lattice. This study naturally acts as a bridge between the idealised two-charge benchmark of Sec.~\ref{sec:method-point-charge-exp} and the DFT data considered later in Sec.~\ref{sec:method-schottky-DFT}. Neutral Sr--O Schottky pairs are created by removing one cation-anion pair from the supercell. The aforementioned integer charges are then assigned to all remaining atoms, and Ewald energies and forces are evaluated analytically following the same procedure as in Sec.~\ref{sec:method-point-charge-exp}. This again leads to exact long-range reference data free of DFT noise, but now representative of an extended medium. The overall supercell remains charge-neutral, but the Sr--O Schottky pair introduces a localised charge distribution, dominated by a large dipole, whose electrostatic interactions the models must learn to reproduce. 

\begin{figure}[h!]
    \centering
    \includegraphics[width=1.0\linewidth]{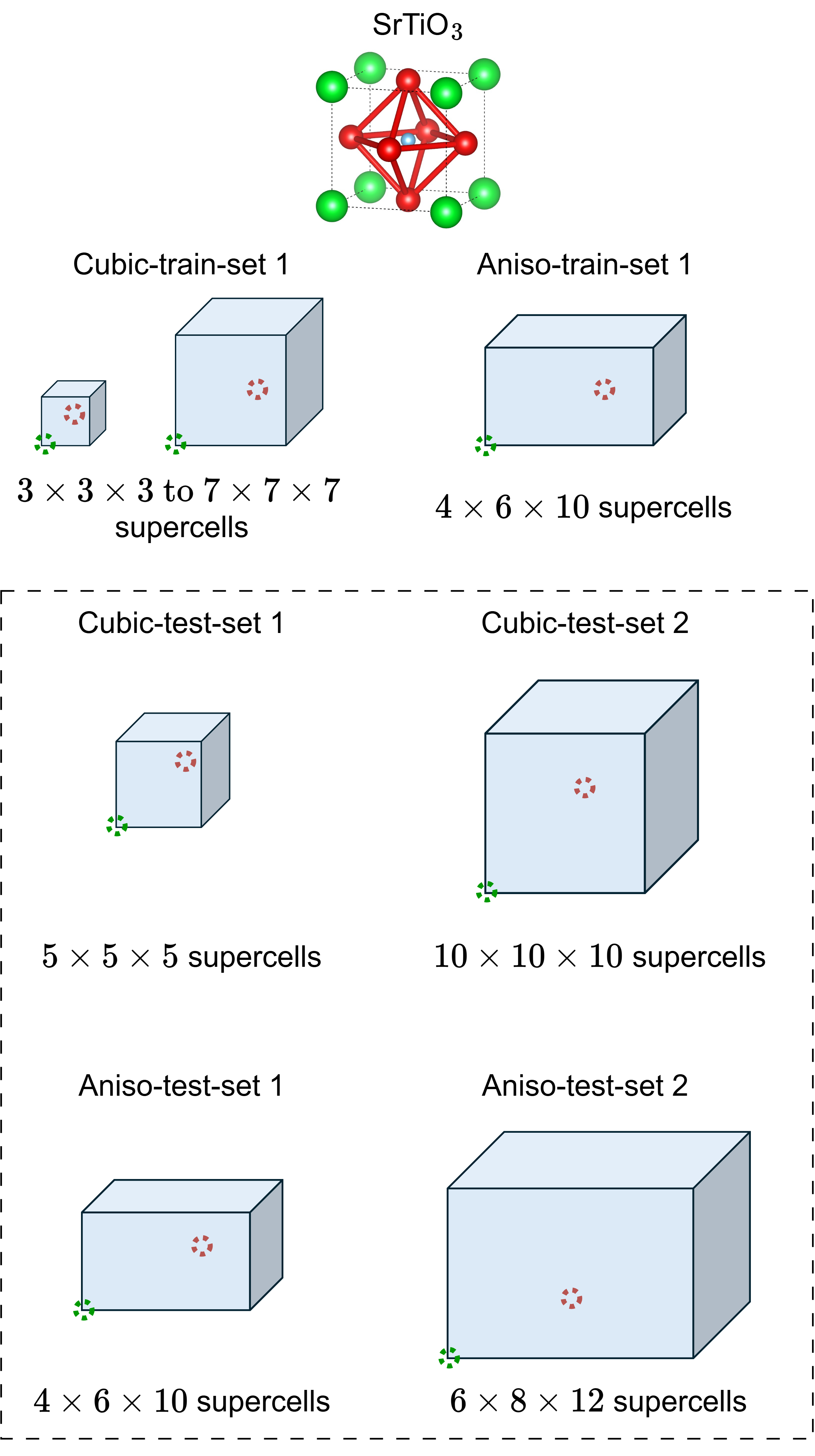}
    \caption{\label{fig:formal_charges_training_set} Illustration of the training and test sets for the mediated Ewald formal-charges SrTiO$_3$ system. Green and red dotted circles represent the Sr and O vacancies respectively, illustrating an example Schottky defect pair. All other atoms in the supercell are held at their formal charge state: O$^{2-}$, Sr$^{2+}$, Ti$^{4+}$.}
\end{figure}

Compared to the point charge system, the configurational space available for training is considerably more constrained. This is because defect separations cannot be varied continuously but are instead determined by the discrete lattice sites within the supercell. To investigate how training set composition affects learning efficiency, we construct two distinct sets of training data, namely the cubic-train-set and aniso-train-set, as illustrated in Fig.~\ref{fig:formal_charges_training_set}. The first set uses an irregular asymmetric supercell created from $4 \times 6 \times 10$ repeats of the primitive unit cell, which provides a large number of inequivalent defect pair separations within a single supercell geometry. The second consists of cubic supercells of increasing size, $3 \times 3 \times 3$, $5 \times 5 \times 5$, and $7 \times 7 \times 7$ supercells, where each size introduces defect pairs at progressively larger separations, directly analogous to the size-transferability tests performed in the point charge experiments. All bulk structures are generated using the Atomic Simulation Environment software (ASE) \cite{hjorth_larsen_atomic_2017}.

For each defect pair configuration across the two training sets, we include both the ideal unperturbed geometry and a randomly rattled variant, in which every Cartesian component of each atomic position is displaced by a random value drawn from a normal distribution with a standard deviation of 0.1 \AA. This breaks the artificial symmetry of the perfect lattice and encourages the model to learn a smoother, more physically representative potential energy surface. 

\subsection{DFT reference data on SrTi$\text{O}_3$} \label{sec:method-schottky-DFT}
The synthetic benchmarks in Sec.~\ref{sec:method-point-charge-exp} and ~\ref{sec:method-schottky-synthetic} provide clean reference data in which the electrostatic signals are simple by construction. However, they do not reflect the complexity of real first-principles calculations, where the total energy contains many competing contributions beyond bare Coulomb interactions. These include exchange-correlation, dispersion energies, and partial screening of the long-range electrostatic signal by electronic redistribution around the defect, which reduces the magnitude of the long-range electrostatic signal relative to the total energy, making it harder for models to isolate and learn. Therefore, moving to DFT reference data leads to more challenging settings and constitutes the most pragmatic test of whether the long-range models can generalise to realistic training conditions. 

All DFT calculations were performed using the all-electron code FHI-aims (version 240507) \cite{blum_ab_2009}, with the Perdew--Burke--Ernzerhof (PBE) generalised gradient approximation as the exchange-correlation functional \cite{perdew_generalized_1996}. A frozen-core approximation was applied throughout, with a cutoff of $-600$ eV to exclude deep core electrons from the variational optimisation in the SCF \cite{yu_accurate_2021}. Given that Ti adopts a single oxidation state of +4 in SrTiO$_3$, spin-polarisation was not included. Brillouin zone sampling and convergence with respect to $k$-point density were verified on bulk SrTiO$_3$ supercells prior to production calculations. Force convergence thresholds for structural relaxation were set to $0.05$\,eV\,\AA$^{-1}$ to maintain computational tractability.  

We assembled a training dataset spanning several categories of SrTiO$_3$ configurations to ensure broad coverage of the relevant configuration space. For bulk structures, we include cubic phase of SrTiO$_3$ sampled across a range of temperatures via molecular dynamics (MD), drawing from structures generated with the MACE-OMAT-0 foundation model \cite{batatia_foundation_2023} and MD trajectories from the NOMAD repository \cite{zacharias_fully_2020} covering a temperature range of $140$--$1600$\,K. Uniaxial strain perturbations of $\pm 10\%$ along one lattice direction were also included to improve coverage of the potential energy surface away from equilibrium. In addition to bulk configurations, we include Schottky defect structures in several supercell geometries, comprising geometry relaxation trajectories and structures pre-relaxed using the MACE-OMAT-0 foundation model \cite{batatia_foundation_2023} for which single-point DFT calculations were subsequently performed. A full breakdown of the dataset composition and key model hyperparameters is given in Table S1 and Table S2 in the Supplemental Material (SM). 

All four models were trained on identical dataset partition, giving each an equal opportunity to learn the underlying potential energy surface. All models achieve comparably low training-set errors (energy RMSE below $1.5$ meV/atom; force RMSE below $20$ meV/\AA), confirming that each architecture was able to fit the training distribution to a similar degree of accuracy. Having established that training was successful and comparable across models, we next examine how well this fit generalises to defect configurations not included in the training set.

\section{Model Performance}

Throughout Secs.~\ref{sec:result-pt_charge_exp} and~\ref{sec:result-schottky-electrostatic-STO}, we report energies relative to the configuration with the shortest point-charge or defect separation, rather than raw absolute energies. This choice is necessary, not merely a matter of presentation. Because each model's absolute energy scale carries a model-dependent additive offset, absolute energies are not directly comparable between models. Referencing every configuration to the shortest-separation case cancels this offset by construction, so that what remains is the separation-dependent interaction energy itself, which is the quantity of physical interest. Accuracy across the full range of separations, not only at long range, is essential. The short-range regime governs the depth of the binding well between the defects. If a model misdescribes this regime, any traps or migration barriers extracted from downstream calculations, such as the NEB analysis in Sec.~\ref{Analysis:NEB}, will be correspondingly unreliable, even if the model's long-range asymptotic behaviour is qualitatively correct. We therefore assess model performance over the entire separation range rather than focusing on the long-range limit alone.

At the end of each results subsection, we summarise model performance in a table using a green/yellow/red scheme. Because each section differs substantially in energy scale, we apply separate thresholds to each. 

For the synthetic point-charge benchmark, which involves only two point charges under idealised electrostatics and therefore yields larger absolute interaction energies, a model is rated green (pass) for relative-energy MAE below 0.1~eV/atom, yellow (partial pass) for errors between 0.1 and 1~eV/atom, and red (fail) for errors of 1~eV/atom or larger. For the formal-charge Ewald and DFT-derived Schottky-pair benchmarks, each model is assessed both numerically and qualitatively, and the worse of the two determines its rating. Numerically, a model is rated green for relative energy MAE below 1~meV/atom, a commonly used accuracy target for MLIPs, yellow for errors between 1 and 10~meV/atom, and red for errors of 10~meV/atom or larger. Qualitatively, a model is rated yellow if it reproduces the overall trend of the reference but shows visible deviations in the shape of the relative-energy curve, such as a shifted minimum or incorrect curvature, and red if it produces the wrong qualitative trend, for instance a flat relative-energy curve where the reference increases monotonically with separation. The qualitative criterion is particularly relevant for the DFT-derived Schottky tests, where the relative energies are diluted over large supercells and small per-atom errors can still correspond to visibly incorrect curves. 

\subsection{Synthetic Unmediated Point Charge System} \label{sec:result-pt_charge_exp}
The first test was performed on the simple two point charge system described in Sec.~\ref{sec:method-point-charge-exp}, with the point charges placed at larger, unseen separation distances (test set 1 from Fig.~\ref{fig:point_charge_dataset}).

\begin{figure}[htbp]     
    \centering
    \includegraphics[width=1.0\linewidth]{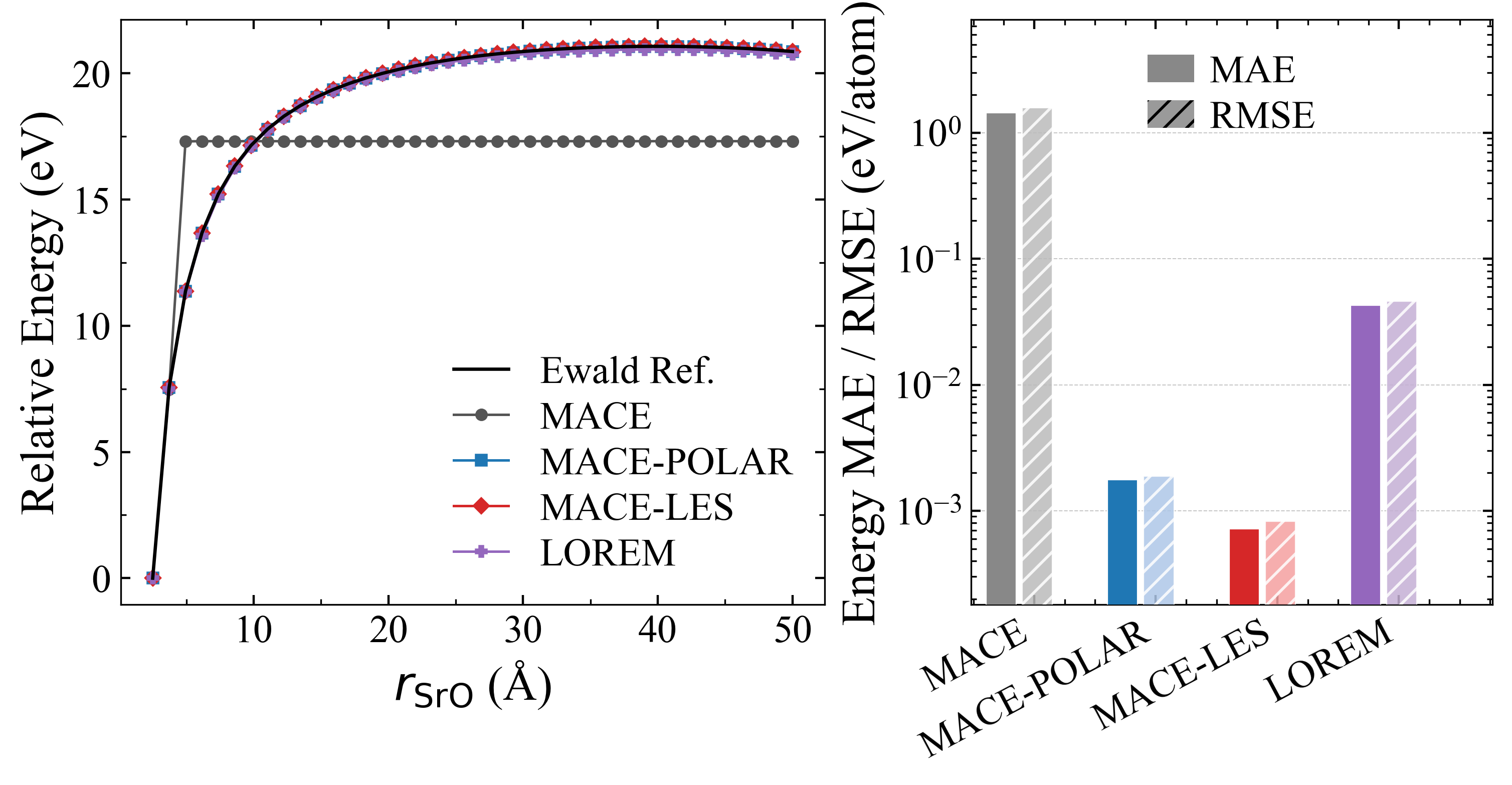}
    \caption{(Left) Relative energies for all four models compared against analytical Ewald reference energies for an $80 \text{\AA} \times 80 \text{\AA} \times 80\text{\AA}$ cell as a function of point-charge separation, zeroed at the shortest separation. The test set explicitly includes unseen distances. (Right) MAE and RMSE in log scale of the relative energies for the four models.}
    \label{fig:point_charge_test_1}
\end{figure}

It is clear from Fig.~\ref{fig:point_charge_test_1} that all long-range models showed promising results in generalising to long-range electrostatics, while short-range MACE fails to capture the correct Coulombic behaviour. The failure of short-range models is expected: the underlying graph neural network performs no message passing beyond the short-range cutoff radius, and therefore has no mechanism to encode interactions at larger separations.  

\begin{figure}[htbp]
    \centering
    \includegraphics[width=1.0\linewidth]{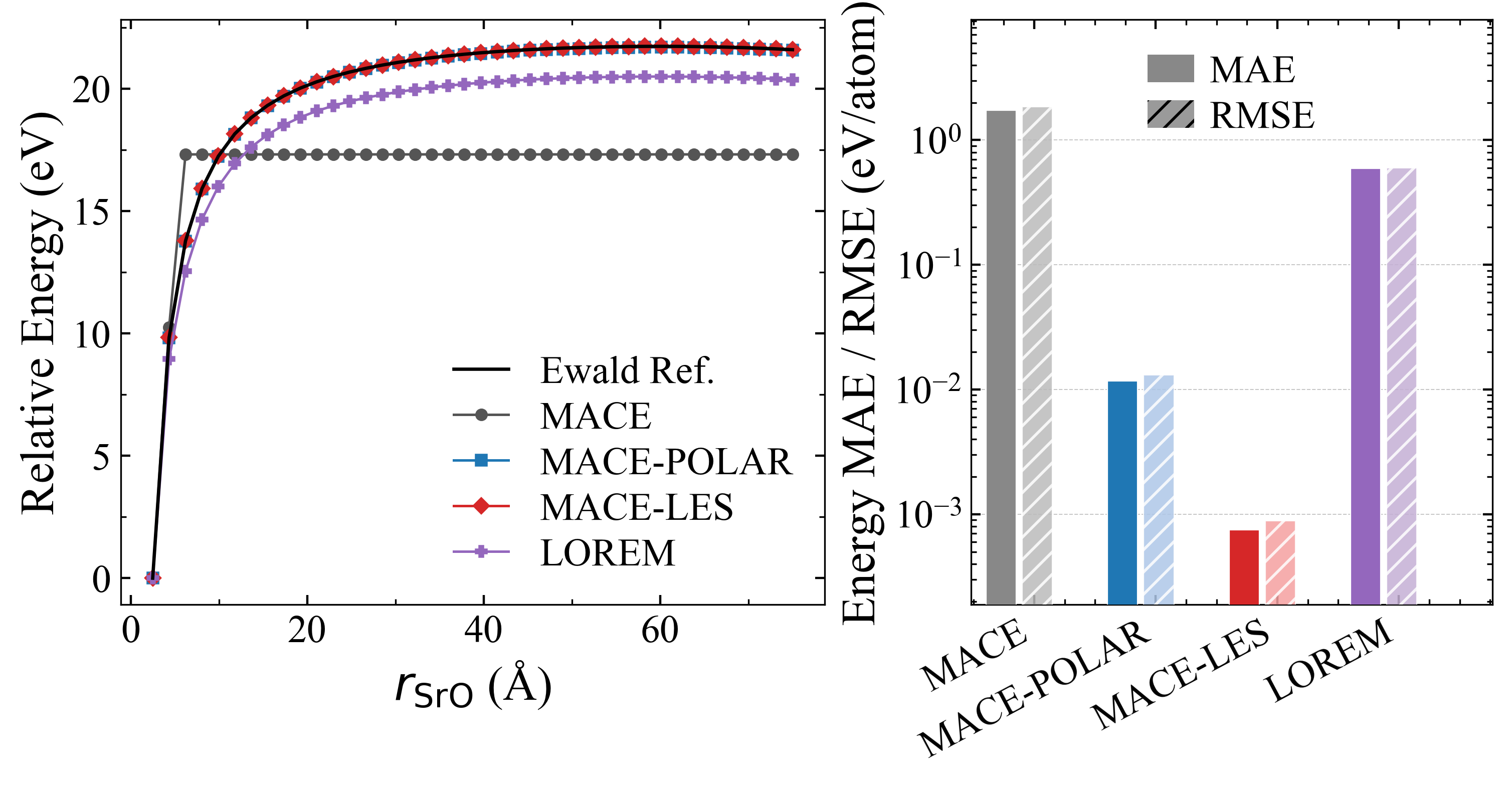}
    \caption{(Left) Relative energies for all four models compared against analytical Ewald reference energies for an  $120 \text{\AA} \times 120 \text{\AA} \times 120\text{\AA}$ cell as a function of point-charge separation, zeroed at the shortest separation. (Right) MAE and RMSE in log scale of the relative energies for the four models.}
    \label{fig:point_charge_test_2}
\end{figure}

Test set 2 from Fig. \ref{fig:point_charge_dataset} examines whether the models generalise to larger simulation cells, a property that is essential for multiscale simulation workflows. As shown in Fig.~\ref{fig:point_charge_test_2}, MACE-POLAR and MACE-LES demonstrated the ability to correctly recover the $\frac{1}{r}$ Coulomb scaling across cell sizes unseen during training, whereas LOREM showed significant deviations from the reference data.  This transferability is a direct consequence of architectural design: both MACE-LES and MACE-POLAR explicitly impose a Coulomb functional form with environment-dependent charges, whose $\frac{1}{r}$ asymptotic are guaranteed by construction and therefore hold for any cell geometry as shown in eq. \ref{eq:mace-les-lr-energy} and \ref{eq:MACE-polar-electrostatics-energy}. In LOREM, by contrast, Ewald summation enters implicitly as a message-passing mechanism that enriches node features; the energy is read out via a subsequent MLP with no explicit $\frac{1}{r}$ constraint enforced on the output. Generalisation to new cell sizes is therefore not structurally guaranteed. 

\begin{figure}[htbp]
    \centering
    \includegraphics[width=1.0\linewidth]{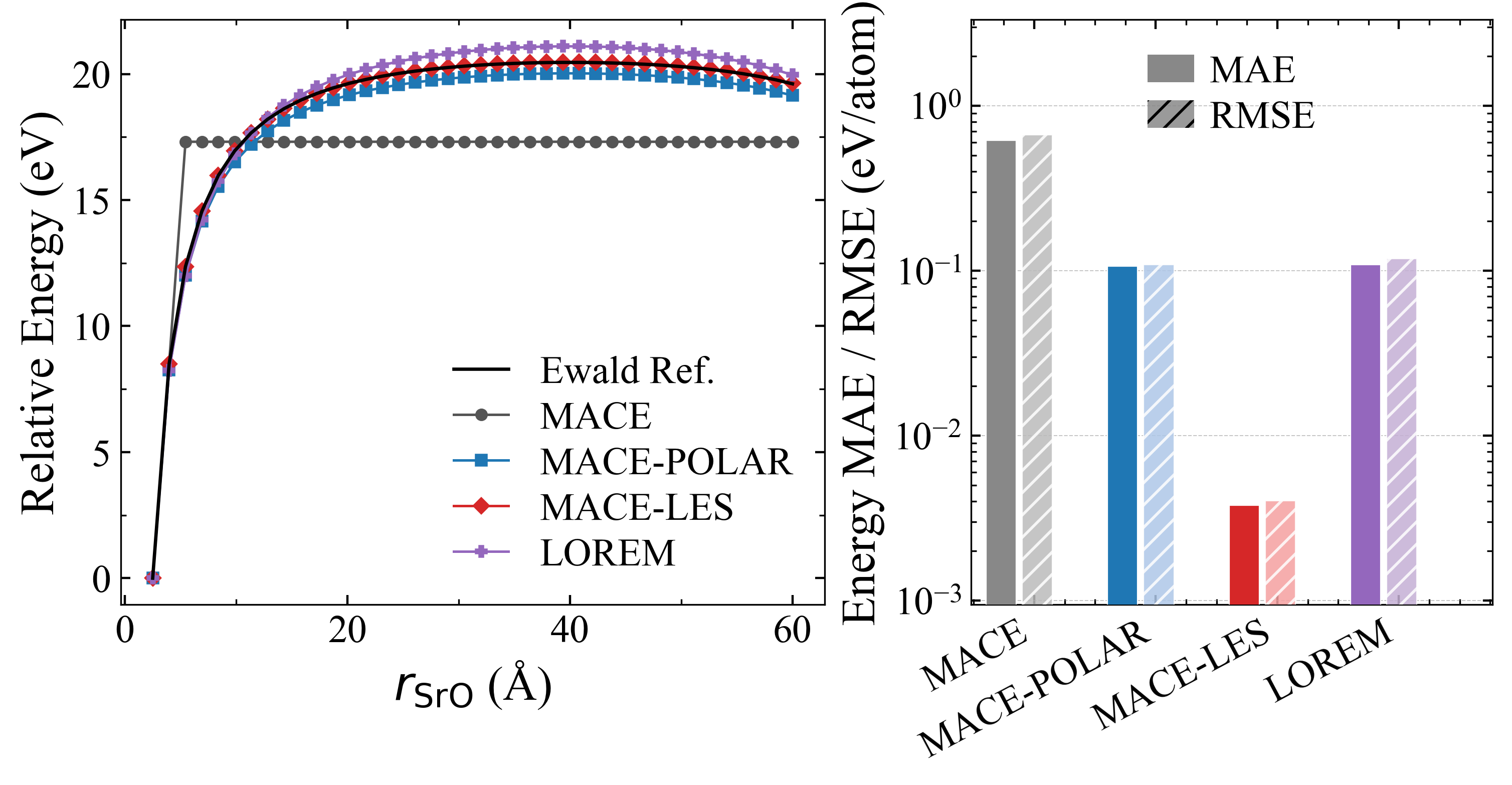}
    \caption{(Left) Relative energies for all four models compared against analytical Ewald reference energies for an $160 \text{\AA} \times 80 \text{\AA} \times 80\text{\AA}$ cell as a function of point-charge separation, zeroed at the shortest separation. (Right) MAE and RMSE in log scale of the relative energies for the four models.}
    \label{fig:point_charge_test_5_phonon}
\end{figure}

Following the unseen distance and large-cell tests, test set 3, a more challenging benchmark illustrated in Fig.~\ref{fig:point_charge_dataset}, was introduced to probe the models' behaviour under conditions relevant to displaced 1D-chain calculations, as described in Sec.~\ref{sec:method-point-charge-exp}. This test implicitly examines whether the model correctly captures the wavevector dependence of the electrostatic interaction, i.e., whether the energy varies correctly as a function of the relative phase of charge displacements across the cell. As shown in Fig.~\ref{fig:point_charge_test_5_phonon}, MACE-LES outperforms all other models on this test. This is attributable to its Ewald-based long-range formalism, which naturally encodes the periodicity of the charge density and is therefore well-suited to capturing long-range interactions under periodic boundary conditions. LOREM underperforms on this test. MACE-POLAR, while noticeably degraded relative to MACE-LES, still performs considerably better than LOREM and MACE, suggesting that its long-range component retains partial sensitivity to the periodic structure.  

\begin{figure}[htbp]
    \centering
    \includegraphics[width=1.0\linewidth]{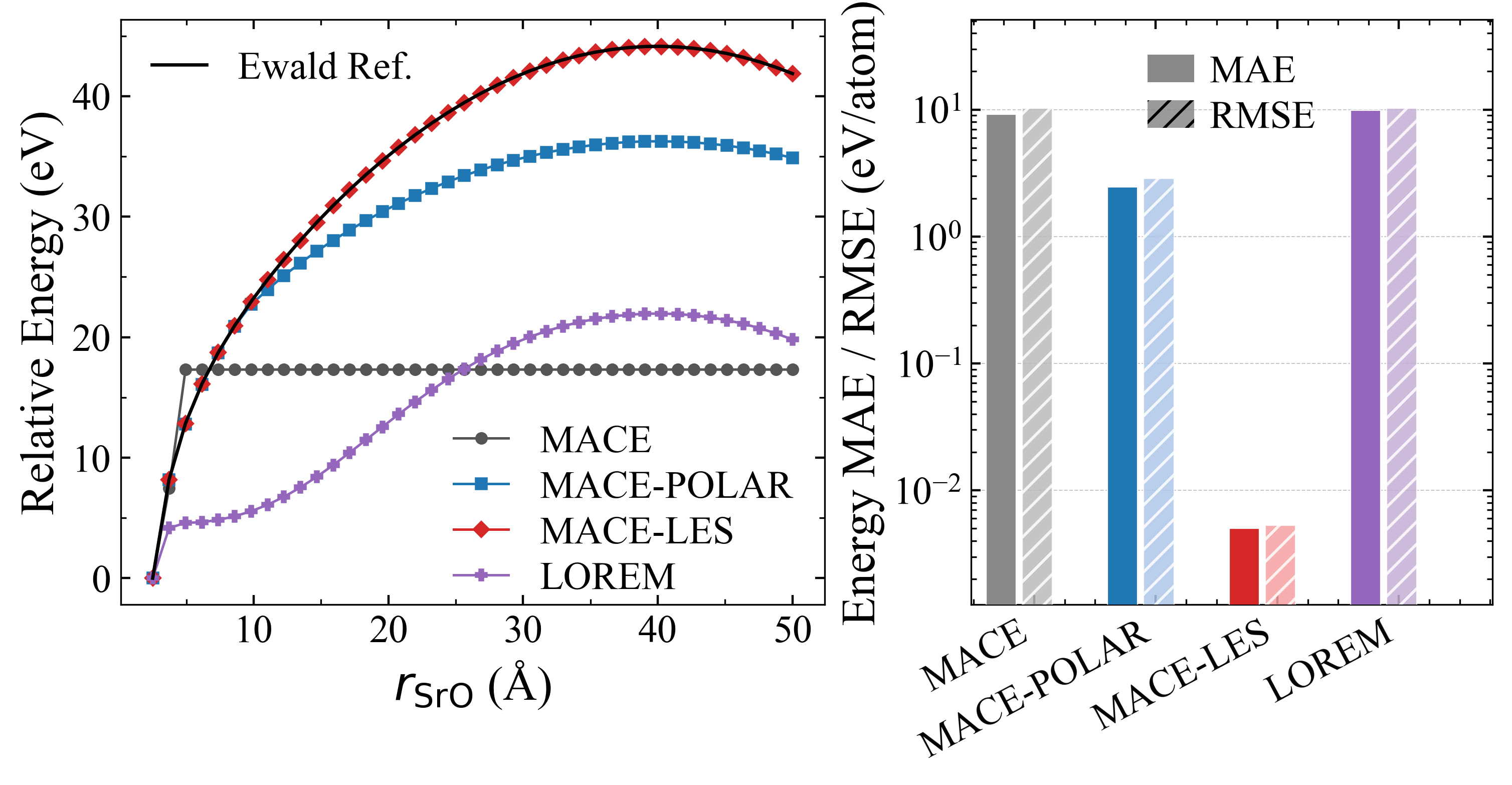}
    \caption{(Left) Relative energies for all four models compared against analytical Ewald reference energies for an $80 \text{\AA} \times 20 \text{\AA} \times 10\text{\AA}$ cell as a function of point-charge separation, zeroed at the shortest separation. (Right) MAE and RMSE in log scale of the relative energies for the four models.}
    \label{fig:point_charge_test_4_anisotropic}
\end{figure}

The results for the final test, test set 4 (Fig.~\ref{fig:point_charge_dataset}), are shown in Fig.~\ref{fig:point_charge_test_4_anisotropic}. This test examines the ability of the models to extrapolate from cubic training cells to an anisotropic cell geometry. MACE-LES performs robustly, smoothly recovering the correct electrostatic behaviour despite the change in cell shape, consistent with its latent charges entering a fixed Coulomb interaction without a learned field dependence. By contrast, MACE-POLAR is less robust to the change in cell shape. Since both models evaluate periodic electrostatics in reciprocal space, and the non-local energy contribution of MACE-POLAR is negligible in this test, the difference is unlikely to arise from the electrostatic energy term; we discuss its likely origin, the flexibility of the learned field-dependent charge updates, in Section \ref{sec:Discussion}. LOREM similarly fails to generalise to the anisotropic geometry. We found that an architectural modification introducing a direct $q_i V_i$ energy skip connection improved performance on the synthetic point-charge benchmarks. However, as this modification degraded performance on the DFT-based benchmarks, the unmodified published model is used throughout; results for this ablation are presented in SM Fig. S1.

Taken together, the four point-charge tests paint a consistent picture as summarised in Table \ref{tab:point_charge_summary}. A pure short-range MACE model cannot reproduce Coulombic interactions beyond the local cutoff at all. LOREM produces long-range interactions via equivariant Ewald message passing, but its capability is not quite robust nor transferable to charge separations and cell sizes unseen during training. Among the models with explicit long-range components, MACE-POLAR performs competitively on simpler generalisation tasks but degrades on the more demanding displaced 1D chain and anisotropic cell tests. MACE-LES, by contrast, demonstrates robust performance across all four tests. Notably, MACE-LES achieves this despite its latent charges not being constrained to physical values. The model learns an effective charge distribution that correctly reproduces the electrostatic behaviour under periodic boundary conditions, suggesting that physical interpretability of intermediate quantities might not be a prerequisite for accurate long-range modelling in this synthetic test case.  

\begin{table}[htbp]
    \centering
    \small
    \caption{Summary of model performance on the synthetic point-charge benchmark tests. $\checkmark$ indicates successful generalisation, $\times$ indicates failure, and $(\checkmark)$ indicates partial success.}
    \label{tab:point_charge_summary}
    \renewcommand{\arraystretch}{1.3}
    \begin{tabular}{lcccc}
        \toprule
            & \textbf{MACE} 
            & \textbf{LOREM} 
            & \textbf{POLAR} 
            & \textbf{LES} \\
        \midrule
        Unseen separation               & \failmark & \passmark    & \passmark    & \passmark \\
        Larger cell                      & \failmark & \partialmark & \passmark    & \passmark \\
        Displaced 1D chain               & \failmark & \partialmark & \partialmark & \passmark \\
        Anisotropic cell                 & \failmark & \failmark    & \failmark    & \passmark \\
        \bottomrule
    \end{tabular}
\end{table}


\subsection{Synthetic Formal Charge Benchmark: Schottky Defects in SrTiO$_3$} \label{sec:result-schottky-electrostatic-STO}

Having established model behaviour on idealised point-charge systems, we turn to SrTiO$_3$ with a Sr--O Schottky vacancy pair, where all remaining atoms retain their formal integer charge states and reference energies are computed via analytical Ewald summation. We report relative energies throughout, removing systematic per-model offsets in atomic energy baselines and isolating each model's ability to capture the variation in interaction energy with pair separation. Models were evaluated across eight test panels shown in Figs.~\ref{fig:STO_ewald_trained_on_cubic_results} and \ref{fig:STO_ewald_trained_on_irreg_results}, covering unseen separations, larger supercells, and rattled configurations, probing both in-distribution generalisation and out-of-distribution extrapolation. We noted that the energies of rattled configurations are not expected to follow a smooth increasing trend with pair separation, as all atomic positions are perturbed randomly, each configuration samples a different point on the potential energy surface, and the vacancy--vacancy interaction is superimposed on this configuration-dependent variation.

\subsubsection{Models trained on cubic-train-set 1}
\begin{figure*}[htbp]
    \centering
    \includegraphics[width=0.9\textwidth]{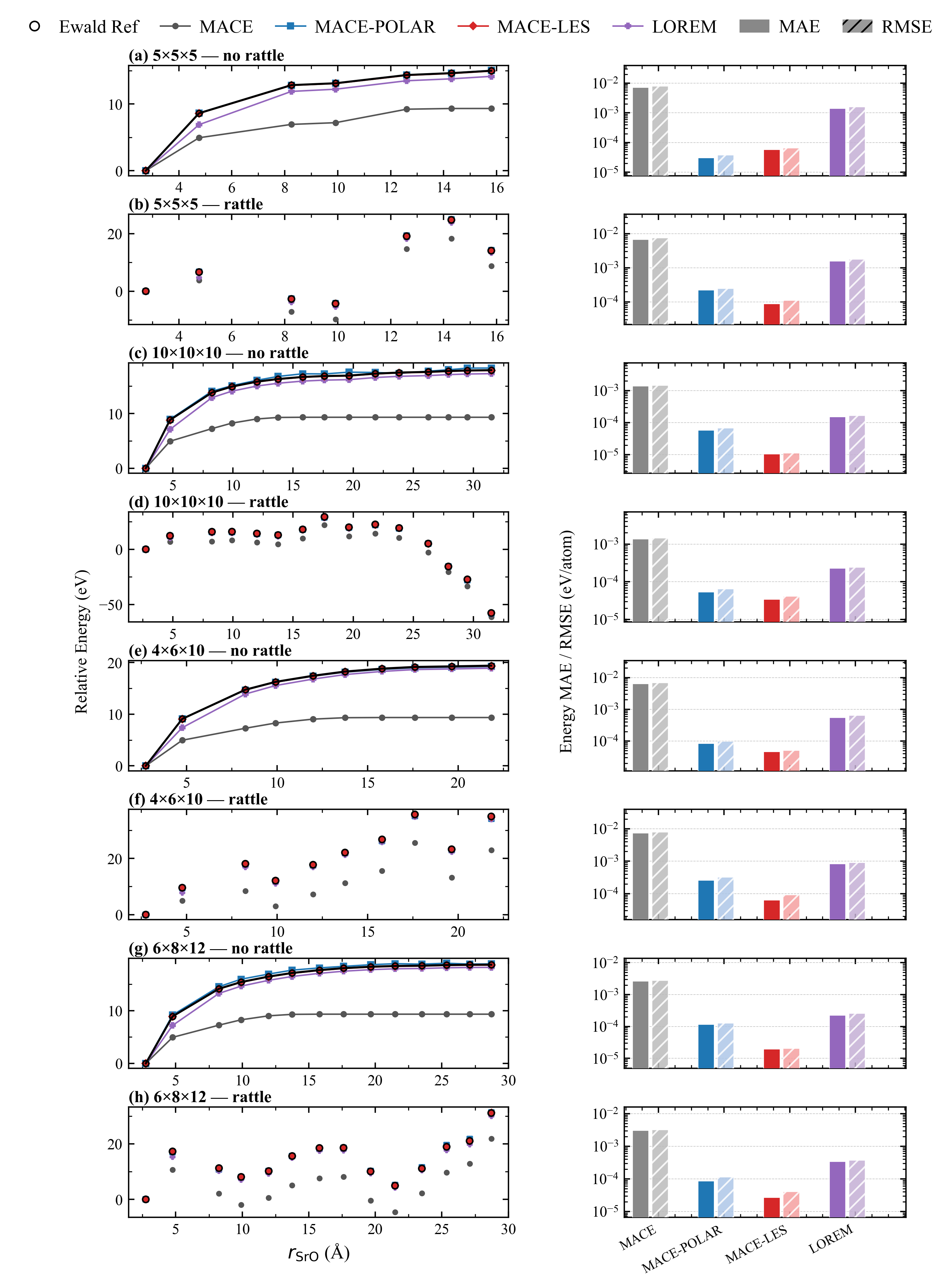}
    \caption{Benchmark results for the four models trained on cubic SrTiO$_3$ with supercell size of $3\times3\times3$ to $7\times7\times7$ (cubic-train-set 1). All panels show energies relative to the first vacancy-separation configuration, zeroed independently for each model, and benchmarked against analytical Ewald energies. MAE and RMSE are reported for each model.
    (\textbf{a}, \textbf{b})~$5\times5\times5$ supercell, unseen separation distances;
    (\textbf{c}, \textbf{d})~$10\times10\times10$ supercell, unseen during training;
    (\textbf{e}, \textbf{f})~$4\times6\times10$ anisotropic supercell, unseen during training;
    (\textbf{g}, \textbf{h})~$6\times8\times12$ anisotropic supercell, unseen during training.
    Panels (a), (c), (e), (g) show ideal atomic positions; panels (b), (d), (f), (h) show rattled atomic positions.}
    \label{fig:STO_ewald_trained_on_cubic_results}
\end{figure*}

As shown in Fig. \ref{fig:STO_ewald_trained_on_cubic_results}, for models trained on cubic supercells, MACE-LES demonstrates the most consistent performance across all eight test panels, correctly capturing the electrostatic behaviour in both rattled and non-rattled configurations and generalising reasonably to anisotropic test geometries. MACE-POLAR performs comparably, mostly achieving slightly higher MAE and RMSE than MACE-LES but remaining competitive across both cubic and anisotropic panels. LOREM ranks similarly to MACE-POLAR but with marginally higher errors, suggesting that its implicit encoding of long-range information via Ewald-based message passing is competitive but does not fully match dedicated long-range architectures. MACE itself shows improvement over its behaviour in the point-charge system: the presence of mediating Ti and O atoms in the crystal structure effectively extends the range of short-range message passing through the graph, partially compensating for the absence of an explicit long-range term. The overall performance is shown in Table \ref{tab:sto_cubic_summary}.

\subsubsection{Models trained on aniso-train-set 1}
\begin{figure*}[htbp]
    \centering
    \includegraphics[width=0.9\textwidth]{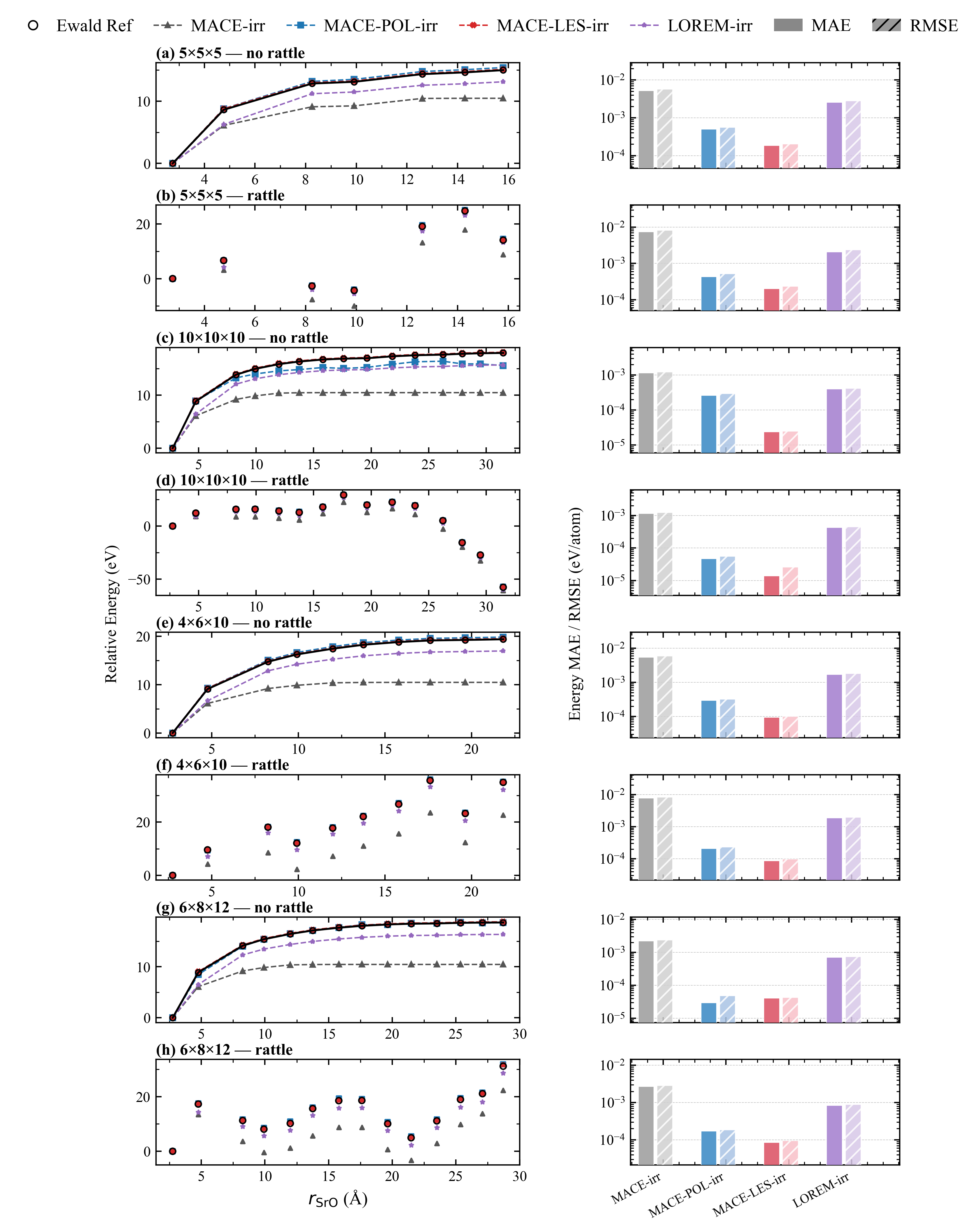}
    \caption{Benchmark results for the four models trained on the anisotropic SrTiO$_3$ $4\times6\times10$ supercell (aniso-train-set 1). All panels show energies relative to the first vacancy-separation configuration, zeroed independently for each model, and benchmarked against analytical Ewald energies. MAE and RMSE are reported for each model.
    (\textbf{a},~\textbf{b})~$5\times5\times5$ cubic supercell, unseen during training;
    (\textbf{c},~\textbf{d})~$10\times10\times10$ cubic supercell, unseen during training;
    (\textbf{e},~\textbf{f})~$4\times6\times10$ anisotropic supercell, seen during training;
    (\textbf{g},~\textbf{h})~$6\times8\times12$ anisotropic supercell, unseen during training.
    Panels (a), (c), (e), (g) show ideal atomic positions; panels (b), (d), (f), (h) show rattled atomic positions.}
    \label{fig:STO_ewald_trained_on_irreg_results}
\end{figure*}

The models trained on anisotropic supercells are named with the suffix ``-irr'', as shown in Fig.~\ref{fig:STO_ewald_trained_on_irreg_results}. Training on anisotropic supercells spanning multiple lattice parameters largely preserves the relative ordering of model performance. From Fig.~\ref{fig:STO_ewald_trained_on_irreg_results}, MACE-LES-irr is the best-performing irregular-trained model in seven of eight tests, achieving the lowest energy MAE on both cubic and anisotropic test geometries. MACE-POLAR-irr remains competitive and ranks second. LOREM-irr performs comparably to MACE-POLAR-irr, with slightly higher errors across panels. MACE-irr continues to occupy the lowest ranking, with the same qualitative failure near the tails. The overall performance is shown in Table \ref{tab:sto_irreg_summary}.

The two training regimes reveal a consistent ranking, with the two models containing an explicit electrostatic energy term, MACE-LES and MACE-POLAR, clearly ahead of LOREM and MACE. Both remain accurate when trained on anisotropic cells, with energy MAEs below 0.2 and 0.5 meV/atom on all test sets for MACE-LES-irr and MACE-POLAR-irr, respectively. MACE-LES is the more robust of the two: its error changes by a consistent factor of two to three between training regimes, whereas the error of MACE-POLAR on the same test set can vary by more than an order of magnitude, in line with its greater flexibility and correspondingly larger data requirements. Notably, both models achieve this accuracy despite the anisotropic training set containing 37\% fewer configurations than the cubic set. Since anisotropic supercells can sometimes be preferred in DFT, both to access larger separations at lower cost and to model systems such as interfaces and grain boundaries, these results indicate that both models, and MACE-LES in particular, are well suited to training data generated in such cells, at least within the range of cell geometries considered here. 

\begin{table}[htbp]
    \centering
    \small
    \caption{Summary of model performance on the SrTiO$_3$ Schottky defect benchmark for models trained on cubic-train-set 1. $\checkmark$ indicates strong performance, $\times$ indicates failure, and$(\checkmark)$ indicates partial success.}
    \label{tab:sto_cubic_summary}
    \renewcommand{\arraystretch}{1.3}
    \begin{tabular}{lccccc}
        \toprule
        \textbf{Test} 
            & \textbf{MACE} 
            & \textbf{LOREM}
            & \textbf{POLAR} 
            & \textbf{LES} \\
        \midrule
        Cubic $5{\times}5{\times}5$, unseen sep. & \failmark & \partialmark & \passmark & \passmark \\
        Cubic $10{\times}10{\times}10$, unseen   & \failmark & \passmark & \passmark & \passmark \\
        Aniso $4{\times}6{\times}10$, unseen     & \failmark & \passmark & \passmark & \passmark \\
        Aniso $6{\times}8{\times}12$, unseen     & \failmark & \passmark & \passmark & \passmark \\
        \bottomrule
    \end{tabular}
\end{table}

\begin{table}[htbp]
    \centering
    \small
    \caption{Summary of model performance on the SrTiO$_3$ Schottky defect benchmark for models trained on aniso-train-set 1. Symbols as in Table~\ref{tab:sto_cubic_summary}.}
    \label{tab:sto_irreg_summary}
    \renewcommand{\arraystretch}{1.3}
    \begin{tabular}{lccccc}
        \toprule
        \textbf{Test} 
            & \textbf{MACE} 
            & \textbf{LOREM}
            & \textbf{POLAR} 
            & \textbf{LES} \\
        \midrule
        Cubic $5{\times}5{\times}5$, unseen            & \failmark & \partialmark & \passmark & \passmark \\
        Cubic $10{\times}10{\times}10$, unseen          & \failmark & \passmark & \passmark & \passmark \\
        Aniso $4{\times}6{\times}10$, in-dist.   & \failmark & \partialmark & \passmark & \passmark \\
        Aniso $6{\times}8{\times}12$, unseen      & \failmark & \passmark & \passmark & \passmark \\
        \bottomrule
    \end{tabular}
\end{table}

\subsection{Real Schottky defect SrTiO$_3$ data} \label{sec:result-schottky-DFT}
As mentioned in Sec. \ref{sec:method-schottky-DFT}, DFT data were used to train bespoke models. The data contains SrTiO$_3$ 1D strain data, bulk MD configurations, and most importantly the Schottky defect datasets. The strain and MD configurations are present to provide a baseline such that the model could gain reasonable knowledge about the species in the bulk material, and obtain correct lattice parameters. We evaluate performance separately on the test sets corresponding to each data category, bulk MD, strain, and Schottky defects, to establish an understanding of the training quality.

\begin{figure}[htbp]
    \centering
    \includegraphics[width=0.5\textwidth]{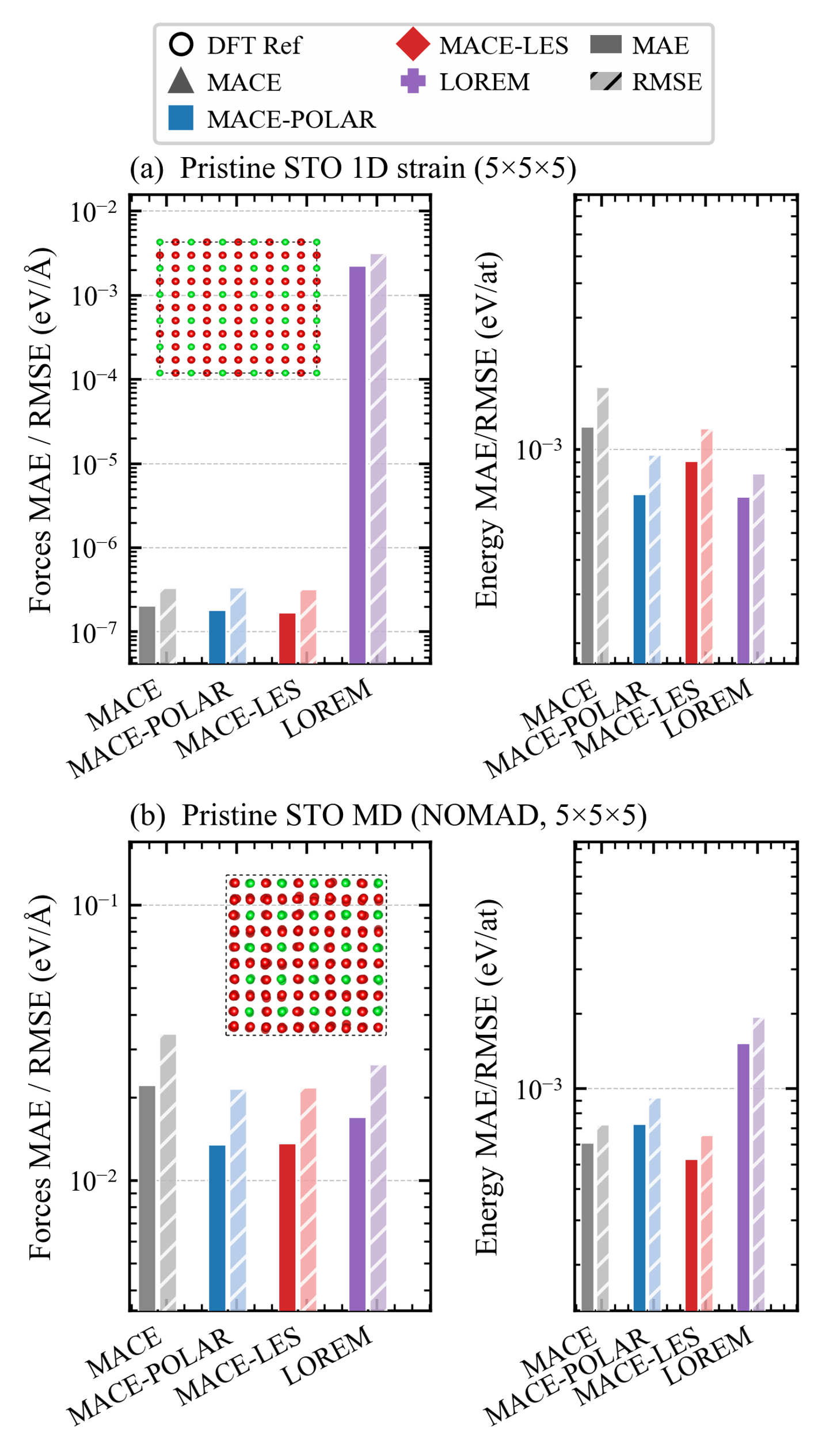}
    \caption{Benchmark of the MLIPs against DFT reference data across two test sets for bulk SrTiO$_3$. The left panel shows forces MAE/RMSE vs.\ Sr–O distance and the right panel shows the energy MAE/RMSE on a log scale. (a) Bulk SrTiO$_3$ under 1D strain ($5\times5\times5$ supercell). (b) Bulk SrTiO$_3$ sampled from NOMAD MD trajectories ($5\times5\times5$ supercell).}
    \label{fig:STO_dft_strain_and_md}
\end{figure}

\begin{figure}[htbp]
    \centering    \includegraphics[width=0.5\textwidth]{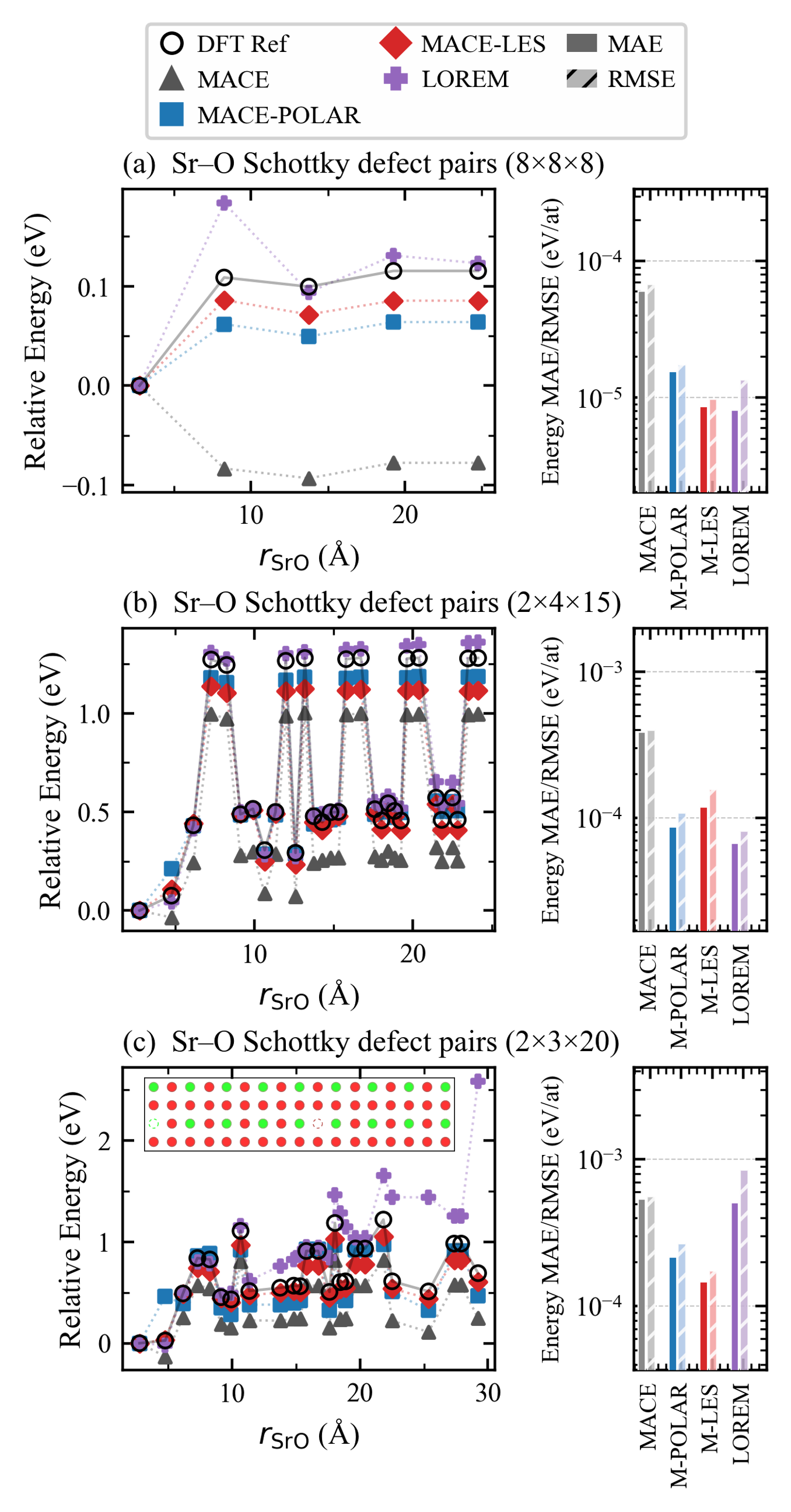}
    \caption{Benchmark of the MLIPs against DFT reference data across the Schottky defect test sets for SrTiO$_3$. The left panel shows the relative energy vs.\ Sr–O separation and the right panel shows the energy MAE/RMSE on a log scale. (a) unseen cubic $8\times8\times8$, (b) seen $2\times4\times15$, and (c) unseen $2\times3\times20 $ SrTiO$_3$ supercells, each containing a pair of Sr–O Schottky defects, with geometries relaxed using the MACE-OMAT foundation model and energies evaluated as single points with DFT and each MLIP.}
    \label{fig:STO_dft_Schottky}
\end{figure}

As seen in Fig. \ref{fig:STO_dft_strain_and_md}, most models capture the energy and force profiles for bulk structures well, while LOREM underperforms somewhat on forces for strained structures and on the relative energy of bulk configurations. This is an important baseline, as regular bulk structures represent the simplest test case and models are expected to handle them reliably. Nonetheless, most models achieve sub-meV accuracy on energy MAE and RMSE (in eV/atom), with force errors around $10^{-2} \mathrm{eV/{\AA}}$. 

Fig.~\ref{fig:STO_dft_Schottky} presents three progressively more demanding test sets: an unseen cubic cell (a), a seen anisotropic cell (b), and an unseen cell with an additional, longer anisotropic lattice vector not present in the training data (c). Across the unseen cubic and seen anisotropic sets, MACE-LES and MACE-POLAR emerge as the strongest performers, achieving comparable errors on energies and forces. In the most demanding extrapolation case, Fig.~\ref{fig:STO_dft_Schottky}(c), MACE-POLAR continues to track the DFT reference closely, and MACE-LES performs comparably well, while LOREM shows a substantial degradation, particularly at the largest Sr--O separations, reflecting the added demands of extrapolating its implicit long-range encoding beyond the training distribution. The short-ranged MACE baseline captures the qualitative trend in relative energy with $r_\mathrm{SrO}$ across all three test sets, but its quantitative agreement with DFT remains substantially worse than any of the long-range models throughout. 

We also note that the defect energies in Fig.~\ref{fig:STO_dft_Schottky} do not vary monotonically with the vacancy pair separation for two reasons. First, the scalar separation does not uniquely determine the electrostatic environment. In an anisotropic supercell, configurations with similar pair distances can differ substantially in their orientation relative to the lattice vectors, and hence in their distances to periodic images. The interaction energy is therefore not a single-valued function of separation, and deviations and jumps are expected. Second, each configuration is fully relaxed, so the total energy includes a configuration-dependent lattice relaxation contribution. The local distortion around each vacancy depends on its site and on the proximity of the partner vacancy, which introduces additional variation on top of the bare electrostatic interaction. The benchmark therefore tests whether models reproduce the DFT energies configuration by configuration, rather than whether the energies follow a smooth distance dependence.  

\begin{table}[htbp]
    \centering
    \caption{Summary of model performance on the SrTiO$_3$ DFT data. Symbols and criteria are as in Table~\ref{tab:sto_cubic_summary}, except for the bulk MD test, which consists of thermally sampled snapshots without a well-defined relative-energy trend and is therefore rated on the force MAE: $\checkmark$ below 20~meV/\AA, $(\checkmark)$ between 20 and 100~meV/\AA, and $\times$ above 100~meV/\AA.}
    \label{tab:sto_dft_summary}
    \renewcommand{\arraystretch}{1.3}
    \begin{tabular}{lcccc}
        \toprule
        \textbf{Test} 
            & \textbf{MACE} 
            & \textbf{LOREM}
            & \textbf{POLAR} 
            & \textbf{LES} \\
        \midrule
        Bulk MD $5\times5\times5$ & \passmark & \passmark & \passmark & \passmark \\
        1D strain $5\times5\times5$ & \passmark & \passmark & \passmark & \passmark \\
        Schottky defect                        & \partialmark & \partialmark & \passmark & \passmark \\
        \bottomrule
    \end{tabular}
\end{table}

\section{Discussion} \label{sec:Discussion}
As anticipated, MACE is a purely short-range semilocal model, and its limitations become pronounced when charged defects are separated beyond its interaction cutoff. In the synthetic point-charge benchmark, this manifests as a characteristic plateau in the predicted Ewald energy: once the charge separation exceeds the cutoff radius, the atoms no longer interact and the model energy saturates, producing the distinctive kink seen in the results. Interestingly, performance improves substantially when moving to the real DFT data, where two effects conspire to extend the model's effective range. Firstly, the presence of intervening atoms mediates longer-range information transfer through the message-passing graph, extending the effective interaction radius beyond its nominal cutoff. Secondly, the high dielectric constant of SrTiO$_3$ screens the bare Coulomb interaction between defects, reducing the spatial range over which the electrostatic signal must be captured. Together, these factors allow MACE to reproduce qualitative energy trends, including relative and formation energies, but the absolute accuracy remains insufficient for quantitative defect modelling.  

MACE-LES emerges as the best-performing model overall in this benchmark. Its long-range modification is conceptually straightforward: a long-range energy term is constructed via Ewald summation over latent charges, which are learned from the training data rather than constrained to reproduce a physical charge distribution. Because these charges enter a fixed Coulomb interaction without any learned dependence on the electrostatic field, the model has little freedom to extrapolate incorrectly, which likely contributes to its consistent performance across cell geometries and training regimes. This also demonstrates that physical interpretability of intermediate quantities is not a prerequisite for accurate long-range modelling, provided the functional form of the long-range interaction is correctly encoded.

MACE-POLAR delivers competitive performance across the DFT-derived benchmarks, with errors close to those of MACE-LES. Its main weakness appears on the synthetic point-charge benchmark under strongly anisotropic cell geometries. We attribute this primarily to the high expressivity and flexibility of its field-dependent charge updates combined with the limited training data. Trained only on cubic cells, these learned updates are poorly constrained in the extreme field environment of the anisotropic cell, so the predicted charges deviate from the correct values. Consistent with this, the learned Fukui functions, which are read out from the same field-dependent features as the charge updates (Eq.~\ref{eq:MACE-polar-Fukui}), deviate increasingly from their values in cubic cells as the anisotropy grows (SM Fig. S2). This weakness does not carry over to the anisotropic cells of the SrTiO$_3$ benchmark (Sec.~\ref{sec:result-schottky-electrostatic-STO}), where MACE-POLAR generalises without breakdown. This is consistent with the larger training sets available there, and possibly also with screening and a more realistic charge density keeping the field environment less extreme than in the idealised point-charge construction. Overall, this flexibility makes MACE-POLAR best suited to large-scale training as a foundation model~\cite{batatia_mace-polar-1_2026}, rather than to training from scratch on small, narrowly sampled datasets. Also, for such synthetic settings, an interesting direction would be a tensorial extension of the Fukui scheme, allowing the charge-deficit correction to be distributed directionally around each atom rather than as a purely spherical monopole. Because the total-charge constraint acts only on the monopoles, total charge and spin would still be conserved. Such a response may improve robustness under extreme electrostatics and limited training data.

LOREM presents a particularly interesting case. It uses latent equivariant charges and Ewald summation to propagate long-range equivariant messages, but these messages only update the node features, meaning there is no explicit electrostatic energy term and no charge conservation. This design results in weaker performance on the synthetic point-charge benchmark, where a physically accurate electrostatic energy is essential, as supported by the improvement obtained with a direct energy term $q_iV_i$ (SM Fig.~S2). However, in some DFT setting, LOREM performs competitively, suggesting that its long-range message passing is well suited to capturing the distributed many-body correlations present in realistic ionic configurations. More generally, the realistic DFT data appear to be a less discriminating test than the idealised point-charge benchmark, where screening, distributed charge densities and other energy contributions likely mask the electrostatics. The relative merit of implicit versus explicit long-range treatments therefore depends on the nature of the target data.

A further limitation for all the models trained in this work is that, while they are capable of accurate total energy differences between configurations, typically they do not reproduce the absolute total energy correctly. Each model exhibits a constant, structure-independent offset relative to the DFT reference. This offset cancels exactly whenever two energies of the same stoichiometry are subtracted, for example, when evaluating formation energy referenced to fixed chemical potentials, so the relative energies used throughout this work remain physically meaningful. However, the offset does not cancel when comparing structures of different stoichiometry, since there is no reason for two compositions to share the same absolute-energy error. We interpret this as potentially connected to the fact that each model is trained jointly on bulk and defect structures without an explicit, charge-neutral reference state, so the implicit chemical potential of each defect species is not constrained during training. A practical consequence is that none of the models tested here can be used to directly compare defects of different stoichiometry: comparing the relative stability of the Sr--O Schottky pair studied in this work against a Ti--O$_2$-type Schottky defect would require the absolute energy of each species to be correct, since this comparison cannot be reduced to a same-stoichiometry energy difference. This question therefore falls outside the scope of the present benchmark and would require either explicit calibration of elemental reference energies for each model or a dedicated training strategy that constrains absolute energies across compositions.

\section{Applications}
Having benchmarked all four models on both synthetic and DFT data, we now apply all the long-range MLIPs together with short-range MACE as baseline, to a physically relevant transport property of SrTiO$_3$: oxygen vacancy migration. This process is of considerable experimental and technological interest, as oxygen ion conductivity underpins the performance of SrTiO$_3$-based memristive and electrochemical devices~\cite{su_oxygen_2024}.  

\subsection{Activation Energy for Oxygen Migration} \label{Analysis:NEB}
\begin{figure}[htbp]
    \centering
    \includegraphics[width=1.0\linewidth]{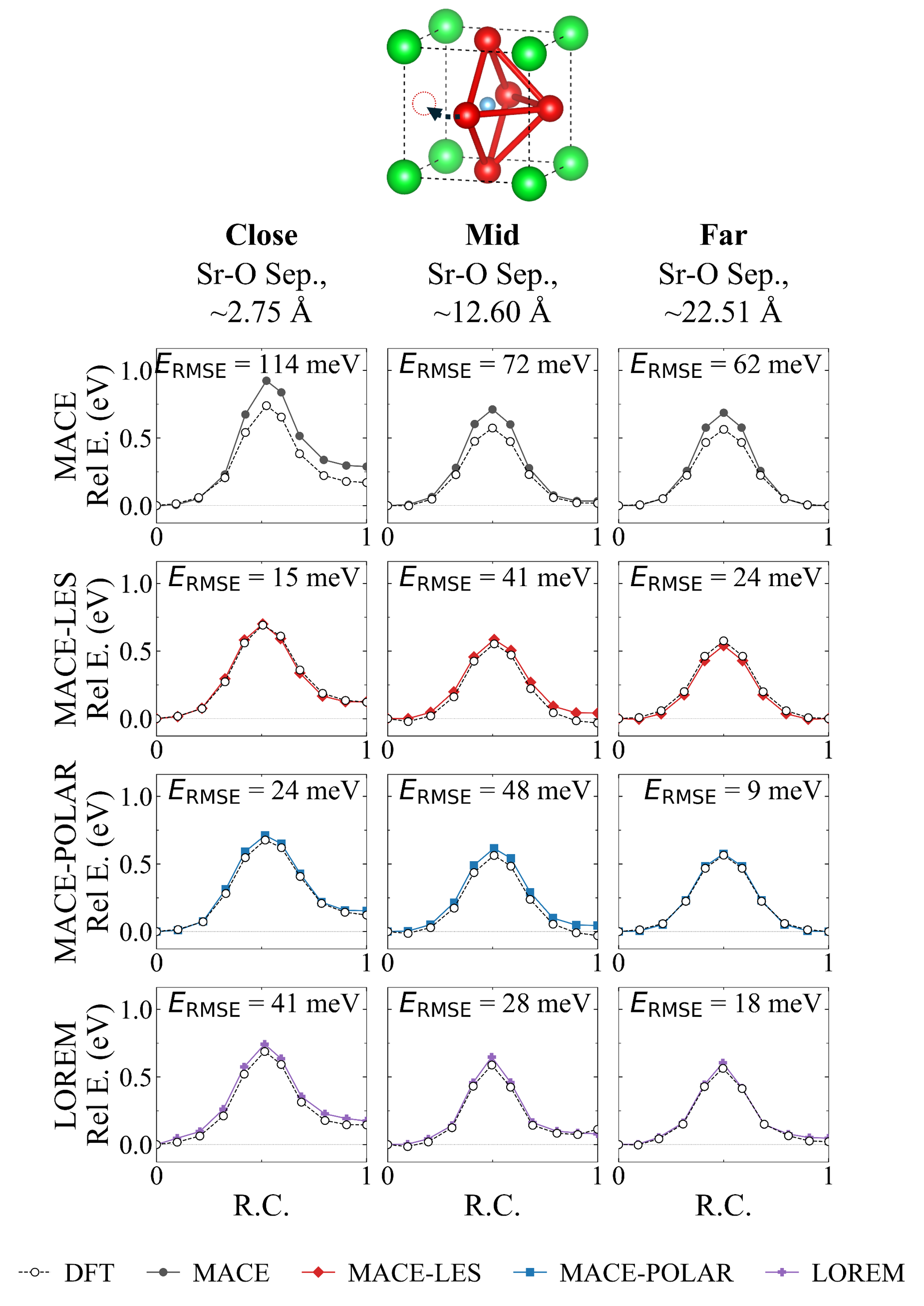}
    \caption{\label{fig:neb_models_1} NEB energy profiles for oxygen vacancy migration in SrTiO$_3$ at close (${\sim}2.75$~\AA), mid (${\sim}12.60$~\AA), and far (${\sim}22.51$~\AA) Sr--O separations. Rows correspond to the four interatomic potentials (MACE, MACE-LES, MACE-POLAR, and LOREM); columns correspond to the three Sr--O separation configurations. Black dashed curves with open circles denote DFT single-point energies evaluated at the ML-relaxed NEB images; coloured solid curves show the corresponding ML predictions. The reaction coordinate (denoted as R.C.) is normalised to the range $[0,1]$, where 0 and 1 denote the start and end of the migration pathway respectively.} 
\end{figure}


\begin{table}[htbp]
\centering
\caption{NEB metrics across three Sr--O separation regimes. $E_\mathrm{RMSE}$ and $F_\mathrm{RMSE}$ measure energy and force errors relative to DFT across all images; $E_\mathrm{A,f}$ and $E_\mathrm{A,b}$ are the forward and backward activation barriers.}
\label{tab:neb_metrics_config1}
\footnotesize
\setlength{\tabcolsep}{4pt}
\begin{tabular}{l r r r r r r}
\toprule
 & $E_\mathrm{RMSE}$ & $F_\mathrm{RMSE}$
 & \multicolumn{2}{c}{$E_\mathrm{A,f}$ (meV)}
 & \multicolumn{2}{c}{$E_\mathrm{A,b}$ (meV)} \\
\cmidrule(lr){4-5}\cmidrule(lr){6-7}
Model/Config & (meV) & (meV/\AA) & DFT & ML & DFT & ML \\
\midrule
\multicolumn{7}{l}{\textit{Close}} \\[\jot]
MACE       & 114.3 & 5.5 & 740 & 924 & 569 & 635 \\
MACE-LES   &  14.8 & 3.9 & 693 & 699 & 571 & 575 \\
MACE-POLAR &  24.4 & 3.6 & 678 & 711 & 555 & 558 \\
LOREM      &  41.1 & 6.1 & 689 & 742 & 544 & 568 \\
\midrule
\multicolumn{7}{l}{\textit{Mid}} \\[\jot]
MACE       &  72.2 & 5.5 & 574 & 712 & 557 & 679 \\
MACE-LES   &  41.4 & 3.4 & 553 & 584 & 586 & 544 \\
MACE-POLAR &  47.7 & 3.3 & 563 & 616 & 594 & 571 \\
LOREM      &  27.6 & 5.9 & 588 & 647 & 475 & 568 \\
\midrule
\multicolumn{7}{l}{\textit{Far}} \\[\jot]
MACE       &  61.6 & 5.3 & 565 & 687 & 564 & 687 \\
MACE-LES   &  23.5 & 2.9 & 575 & 540 & 576 & 540 \\
MACE-POLAR &   8.9 & 2.8 & 565 & 575 & 565 & 575 \\
LOREM      &  17.8 & 5.9 & 564 & 604 & 542 & 558 \\
\bottomrule
\end{tabular}
\end{table}

Activation barriers for oxygen vacancy migration were computed using the climbing image nudged elastic band (CI-NEB) method, with ML-relaxed images and DFT single-point energies evaluated at the converged geometries to provide a reference. Three Sr--O Schottky pair separations were considered: close (${\sim}2.75\mathrm{\AA}$), medium (${\sim}12.6\mathrm{\AA}$), and far (${\sim}22.5\mathrm{\AA}$), spanning configurations in which the electrostatic interaction between the vacancy pair and the periodic images varies substantially. This range is particularly demanding for short-ranged models, since the Madelung-like contribution to the barrier from charged defects and their periodic images cannot be captured without an explicit long-range term.  

All models reproduce a similar trend, with the activation barrier decreasing as the separation between the two vacancies increases, as shown in Fig.~\ref{fig:neb_models_1}. The asymmetry between the forward and backward barriers arises because the migrating oxygen ion moves between two inequivalent sites: the initial and final configurations of each migration path are not geometrically equivalent. This distinction carries physical information beyond a single scalar barrier height, since the forward and backward processes probe different local environments, which may be relevant when considering percolation pathways for vacancy migration. Table~\ref{tab:neb_metrics_config1} therefore reports both quantities separately: the lower of the forward $E_{A_f}$ and backward barriers $E_{A_b}$, governs the rate of local hopping between two vacancy sites, while the higher of the two acts as the rate-limiting step for longer-range migration, relevant to percolation behaviour. The relative magnitude of $E_{A,f}$ and $E_{A,b}$ therefore encodes information beyond either barrier alone, and their ordering need not be the same at every separation or configuration.  

All three long-range models substantially outperform the short-range MACE baseline in reproducing the DFT reference profiles. The forward and backward barriers are reported in Table~\ref{tab:neb_metrics_config1}, consistent with the experimental and computational literature~\cite{dawson_dynamical_2020, de_souza_behavior_2012}. This performance gain is attributable to the models' treatment of long-range electrostatic interactions between the charged vacancy sites and their periodic images. MACE is unable to reproduce this modulation well, with barrier errors of up to $\sim$0.18~eV relative to DFT, and a correspondingly higher energy RMSE across all three separations, confirming that long-range electrostatics is the decisive physical ingredient for accurate NEB profiles in this system.  

Nonetheless, a distinction emerges among the three long-range models. MACE-POLAR provides the most consistently accurate barriers across all three separation regimes, closely matched by MACE-LES at close and mid separation. LOREM shows a consistent degradation in barrier accuracy relative to MACE-LES and MACE-POLAR at close and mid separation, most pronounced at mid separation, where its backward barrier error (93~meV) is the largest observed among the long-range models in this study. However, at far separation, LOREM's barrier accuracy is comparable to or better than MACE-LES, echoing the trend seen in the aggregate $E_\mathrm{RMSE}$ at mid separation, where LOREM also achieves the lowest error among the three long-range models despite its larger barrier-specific errors there. This suggests that LOREM's error does not scale uniformly with vacancy-pair separation, and may instead be sensitive to the local perturbation of the charge density near the saddle-point configuration itself. MACE-POLAR and MACE-LES are therefore identified as the most reliable models overall for activation barrier prediction in this system, though LOREM remains substantially more accurate than the short-ranged baseline throughout. The energy and force RMSE relative to DFT for all models are reported in Table~\ref{tab:neb_metrics_config1}.

\subsection{Diffusivity from Mean Squared Displacement (MSD)} \label{Analysis:diffusivity}

\begin{figure}[H]
    \centering
    \includegraphics[width=1.0\linewidth]{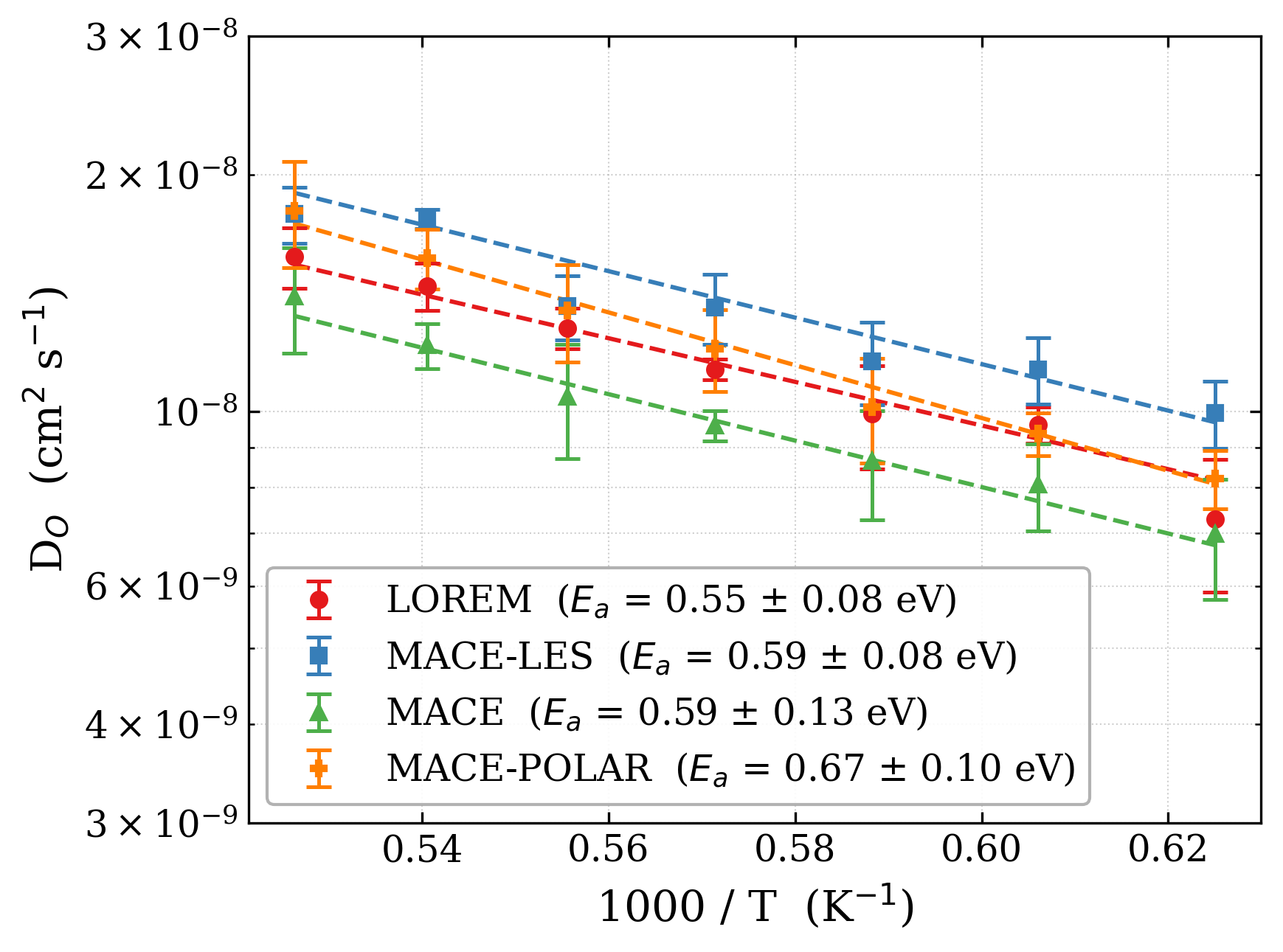}
    \caption{ \label{fig:Arrhenius_plot_DO} Arrhenius plot of the oxygen vacancy diffusion coefficient $D_\mathrm{O}$ as a function of inverse temperature, evaluated between 1600~K and 1900~K in 50~K steps for MACE, MACE-LES, MACE-POALR, and LOREM. Each data point is extracted from the gradient of the linear (diffusive) regime of the MSD, computed from a 2000~ps trajectory (first 20~ps discarded) in a $7{\times}7{\times}7$ supercell (1713 atoms) with a 2~fs Langevin timestep ($\gamma = 0.01$~fs$^{-1}$). Dashed lines show linear fits to $\ln D_\mathrm{O}$ versus $1/T$, with extracted activation energies $E_a$ reported in the legend.}
\end{figure}

Using the Einstein relation for random-walk diffusion \cite{einstein_uber_1905},
\begin{equation}
    D = \lim_{t \to \infty} \frac{\langle r^2(t) \rangle}{2d\,t},
\end{equation}
where $d = 3$ is the dimensionality, oxygen vacancy diffusion coefficients $D_\mathrm{O}$ were extracted from MD trajectories in the NVT ensemble at temperatures between 1600~K and 1900~K in 50~K increments. A $7{\times}7{\times}7$ supercell (1713 atoms) was used throughout to minimise finite-size effects on vacancy migration. Each trajectory was integrated with a 2~fs timestep using a Langevin thermostat ($\gamma = 0.01$~fs$^{-1}$); the first 20~ps were discarded as equilibration of the thermostat, and the MSD was accumulated over the remaining 180~ps of production. Diffusion coefficients were extracted from the gradient of the linear regime of the MSD and plotted as a function of inverse temperature to construct the Arrhenius diagram shown in Fig.~\ref{fig:Arrhenius_plot_DO}. The temperature dependence of $D_\mathrm{O}$ follows the Arrhenius form
\begin{equation}
    D_\mathrm{O}(T) = D_0 \exp\!\left(-\frac{E_a}{k_\mathrm{B} T}\right).
\end{equation}

From this equation the migration activation energy $E_a$ is obtained as the slope of a weighted least-squares fit to $\ln D_\mathrm{O}$ versus $1/T$. MACE yields $E_a = 0.59 \pm 0.13$~eV, MACE-LES yields $E_a = 0.59 \pm 0.08$~eV, MACE-POLAR yields $E_a = 0.67 \pm 0.10$~eV, and LOREM yields $E_a = 0.55 \pm 0.08$~eV. All four values are reasonably consistent within their uncertainties and broadly consistent with the activation barriers obtained independently from the NEB calculations in the preceding subsection. MACE-POLAR's estimate is slightly higher than the other three models, though its uncertainty range overlaps with them, and is also in line with its NEB-derived activation barriers. Some of the spread across all four estimates may reflect statistical limitations inherent to the Arrhenius fitting procedure rather than a genuine difference in the physics. As the diffusive hops become increasingly rare at lower temperatures (see SM Fig. S3 for example MSD trajectories), so the resulting statistics are noisier for a fixed trajectory length at these points. Both longer MD runs and additional independent seeds at the lowest temperatures could potentially help tighten the uncertainty.

\subsection{Comparison with Literature and Scope of Validation}
Direct quantitative comparison of our computed activation barriers and diffusion coefficients with the broader literature requires care, as the majority of prior computational studies consider an isolated oxygen vacancy $V_\mathrm{O}^{\bullet\bullet}$ in a charge-compensated supercell, rather than charge-neutral $\{V_{\mathrm{Sr}}^{''} + V_\mathrm{O}^{\bullet\bullet}\}$ Schottky pairs examined here. First-principles DFT calculations yield migration barriers in the range of 0.3--0.89~eV, depending on supercell size, exchange-correlation functional, and vacancy charge state~\cite{lontsi-fomena_density_2008, carrasco_first-principles_2006, cuong_oxygen_2007, al-hamadany_impact_2013, zhang_oxygen_2016, walsh_strontium_2011}. Classical MD studies using empirical interatomic potentials, which can access much larger supercells, report barriers spanning a comparably wide range of 0.5--0.93~eV~\cite{dawson_dynamical_2020, schie_molecular_2012, waldow_computational_2016}; this spread is attributable primarily to the choice of classical potential rather than to vacancy concentration alone, since differing charge-compensation schemes and functional forms (e.g. Buckingham versus Lennard-Jones or Morse terms) are found to shift the barrier by several tenths of an eV even at comparable, dilute vacancy concentrations~\cite{waldow_computational_2016}. The corresponding oxygen diffusion coefficients reported from these MD studies, typically $D_\mathrm{O} \sim 10^{-10} - 10^{-7} \mathrm{cm^{2}s^{-1}}$ over the simulated temperature range ($900-2000$~K), are generally within one order of magnitude of experimental tracer-diffusion values~\cite{de_souza_behavior_2012}, providing an additional, independent point of comparison alongside the activation barriers. Experimentally, numerous studies over the past several decades have probed oxygen migration in SrTiO$_3$ via tracer diffusion, oxidation kinetics, isotope exchange, transient electrical conductivity, inelastic relaxation, and thermally stimulated depolarization current measurements, reporting activation enthalpies spanning 0.22--1.13~eV on nominally undoped or weakly doped single-crystal samples~\cite{de_souza_behavior_2012, paladino_oxidation_1965, walters_diffusion_1967, yamaji_oxygen-ion_1975, schwarz_determination_1975, liu_thermally_2008, cordero_hopping_2007}. This spread reflects genuine methodological differences, including small acceptor-dopant concentrations in some samples, varying oxygen-vacancy concentrations across measurements, and whether the technique probes isolated-vacancy hopping or vacancy-pair/cluster reorientation. Our ML-NEB value of 0.5--0.7~eV for the far-separation regime is consistent with both ranges towards the lower end.  

To our knowledge, no prior DFT-NEB or MLIP study has systematically examined $V_\mathrm{O}^{\bullet\bullet}$ migration barriers in the presence of a charge-compensating $V_{\mathrm{Sr}}^{''}$ in bulk SrTiO$_3$.  The closest analogue in the literature is the NEB study of Walsh et al.~\cite{walsh_strontium_2011}, who computed O-ion migration barriers both in isolation and adjacent to a strontium vacancy. They found that proximity to a $V_{\mathrm{Sr}}^{''}$ increases the O migration barrier from 0.53~eV to 0.89~eV, an increase of approximately 0.36~eV, which they attribute to the retarding effect of the local electrostatic perturbation introduced by the cation vacancy. This result demonstrates that the migration barrier is sensitive to the charge state and proximity of a neighbouring defect, even though their study considered only a single, isolated Sr--O pair rather than a charge-neutral Schottky pair across a range of separations. The present benchmark directly addresses this gap by testing whether long-range MLIP architectures can faithfully represent the coupled electrostatic landscape of a charge-neutral Schottky pair, which is a prerequisite for any future study of defect-assisted oxygen transport in donor- or acceptor-doped perovskites.

\section{Conclusion} \label{sec:conclusion}
Overall, we have presented a systematic benchmark of long-range machine learning interatomic potentials, namely MACE-LES, MACE-POLAR, and LOREM, progressing from idealised synthetic point-charge systems to DFT data for strontium titanate with Sr--O Schottky vacancy pairs.  

In the synthetic benchmark, MACE-LES emerged as the strongest performer, demonstrating that explicit long-range energy evaluation in reciprocal space confers a meaningful advantage when the interaction is purely electrostatic. In the DFT data benchmark, all three long-range models showed strong generalisation across training regimes, producing correctly shaped and quantitatively reliable minimum energy pathways for oxygen vacancy migration, yielding activation barriers in good agreement with literature values. This is a notable extrapolation, as the training data comprised only bulk MD configurations and relaxed (endpoint) Schottky defect structures, with no intermediate migration geometries resembling the transition-state configurations sampled along the NEB path. By contrast, purely short-range models such as standard MACE remain fundamentally limited by their real-space cutoff, and this limitation becomes especially acute in defected structures where charged defects introduce slowly decaying electrostatic interactions.

However, there exist several important caveats with these findings. The synthetic benchmark, while instructive for isolating electrostatic behaviour, is by design an extreme test case: few real materials present interactions as purely long-range as formal point charges, and a model's underperformance on this task should not be taken as disqualification for most applications. Similarly, the DFT benchmark metrics should not be interpreted as straightforward improvements over the synthetic case. The complexity of the DFT total energy, encompassing short-range repulsion, exchange-correlation, dispersion, and ionic self-energies, dilutes the electrostatic signal and could reduce the associated errors with electrostatics.  

Looking forward, several extensions of this work can be considered. Phonon calculations and the evaluation of finite-temperature thermodynamic properties would test whether the long-range accuracy demonstrated in static benchmarks carries through to dynamical quantities. More complex defect structures, including higher vacancy concentrations and non-neutral supercells with explicit compensating charges, would probe the models' robustness under conditions directly relevant to electrochemical applications such as solid-state batteries, where the repeated insertion and removal of ions leaves residual charge in the electrode material. Finally, a broader survey of strongly polar and ionic materials beyond SrTiO$_3$ would establish the degree to which these conclusions generalise across the wider space of oxide and halide compounds where long-range electrostatics plays a central role.

\begin{acknowledgments}
YCW acknowledges funding from the EPSRC CDT in Modelling of Heterogeneous Systems funded by EP/S022848/1. RJM acknowledges financial support by the Alexander-von-Humboldt foundation. Computing facilities were provided by the Scientific Computing Research Technology Platform of the University of Warwick using the High-Performance Computing (HPC) cluster Avon, and the Sulis Tier 2 platforms at HPC Midlands+ funded by the Engineering and Physical Sciences Research Council (EPSRC), grant number EP/T022108/1. YCW and NDMH acknowledge the computational resources from the UK national high performance computing service, ARCHER2, for which access was obtained via the UK Car-Parrinello Consortium (EPSRC Grant No. EP/X035891/1). YCW and NDMH also acknowledge the use of resources provided by the Isambard-AI National AI Research Resource (AIRR) through project AI4S-01. Isambard-AI is operated by the University of Bristol and is funded by the UK Government's Department for Science, Innovation and Technology (DSIT). MFL acknowledges funding from the German Research Foundation (DFG), project number 544947822.
    
\end{acknowledgments}

\section{Data Availability and Model Versions}
All data supporting the findings of this study will be archived permanently upon acceptance, and are currently available at \url{https://github.com/nickhine/lr_mlip_benchmark}. To allow reproduction, we report the git commit of every research code used, since several were installed directly from source repositories rather than from tagged releases. The MACE-based models (MACE, MACE-LES and MACE-POLAR) were trained and evaluated with \texttt{mace-torch} v0.3.15 (\url{https://github.com/ACEsuit/mace}, commit \texttt{667eee4}, 1~March 2026), built on PyTorch~2.10.0 (CUDA~12.6), \texttt{e3nn}~0.4.4 and ASE~3.27.0. The long-range electrostatic contribution of PolarMACE is provided by \texttt{graph\_longrange} v0.4.0 (\url{https://github.com/WillBaldwin0/graph_electrostatics}, commit \texttt{0135036}, 26~February 2026), and that of MACE+LES by the Latent Ewald Summation package \texttt{les} v0.1.0 (\url{https://github.com/ChengUCB/les}, commit \texttt{18326dd}, 1~April 2026). The LOREM model was trained and evaluated with \texttt{lorem-jax} v0.1.dev17+g95097b38f (\url{https://github.com/lab-cosmo/lorem-jax}, branch \texttt{lr-skip-connection}, commit \texttt{95097b38}, 27~March 2026), using JAX~0.9.2, \texttt{flax}~0.12.6, \texttt{e3x}~1.0.2, \texttt{marathon-train}~0.2.2 and ASE~3.28.0. Its Ewald summation is evaluated with \texttt{jax-pme} v0.1.0a1 (\url{https://github.com/lab-cosmo/jax-pme}, commit \texttt{ebf8d8b4}). Complete frozen environment specifications for both software stacks are distributed with the code repository.

\bibliographystyle{vancouver}
\bibliography{references}

\nocite{*}
\end{document}